\documentclass[superscriptaddress,showpacs,twocolumn,pre]{revtex4-1}
\usepackage[utf8]{inputenc}
\usepackage{graphicx}
\usepackage{color}
\usepackage{appendix}
\usepackage{amsmath}
\usepackage{amssymb}
\usepackage{subfigure}
\usepackage{epsfig}
\newcommand{\kapp}{\boldsymbol{\kappa}}
\newcommand{\pdf}{\Psi(\{\textbf{r}_i\},t)}
\newcommand{\rr}{\textbf{r}}
\newcommand{\om}{\hat{\Omega}}
\newcommand{\omadj}{\hat{\Omega}^{\dagger}}
\newcommand{\F}{\textbf{F}}
\newcommand{\pdfe}{\Psi_{\text{eq}}(\{\textbf{r}_i\})}
\newcommand{\sigg}{\boldsymbol{\sigma}}
\newcommand{\siggh}{\hat{\boldsymbol{\sigma}}}
\newcommand{\B}{\textbf{B}}
\newcommand{\kk}{\textbf{k}}
\newcommand{\qq}{\textbf{q}}
\newcommand{\pp}{\textbf{p}}
\newcommand{\E}{\textbf{E}}
\newcommand{\gam}{\dot{\gamma}}
\newcommand{\strain}{\boldsymbol{\varepsilon}}
\newcommand{\joe}{\color{black} }

\begin{document}
\title{
Normal-stress coefficients and rod climbing in colloidal dispersions
}

\date{\today}
\author{T.F.F. Farage}
%\email{thomas.farage@unifr.ch}
\affiliation{Department of Physics, University of Fribourg, CH-1700 Fribourg, Switzerland}
\author{J. Reinhardt}
%\email{johannes.reinhardt@unifr.ch}
\author{J.M. Brader}
\email{joseph.brader@unifr.ch}
\begin{abstract}
We calculate tractable microscopic expressions for the low-shear normal-stress coefficients of 
colloidal dispersions. 
Although restricted to the low rate regime, the presented formulas are valid for all volume
fractions below the glass transition and for any interaction potential. 
Numerical results are presented for a system of colloids 
interacting via a hard-core attractive
Yukawa potential, for which we explore the interplay between attraction strength and
volume fraction. 
We show that the normal-stress coefficients exhibit nontrivial features close to the critical point
and 
at high volume fractions in the vicinity of the reentrant glass transition.
Finally, we exploit our formulas to make predictions about rod-climbing effects in attractive
colloidal dispersions.
\end{abstract}

\pacs{83.60.Hc, 83.80.Hj, 64.70.pv, 83.10.Gr}
\keywords{Colloid, Nonequilibrium, Rheology}

\maketitle
\section{\label{sec:intro}Introduction}
Complex fluids, {\joe such as colloidal dispersions,} exhibit a nontrivial response
when submitted to an externally applied flow. Depending on the {\joe thermodynamic state point, 
the strain, and the strain rate, nonlinear changes in macroscopic quantities may be observed 
(e.g., thinning or thickening of the shear viscosity \cite{larson1999,wagner_book})}.
In contrast to Newtonian fluids, complex fluids typically exhibit nonzero values of the  
first and second normal-stress differences. 
These rheological functions are of a higher order than the familiar shear viscosity, in the sense
that 
their lowest-order contribution to the flow response is quadratic in the shear rate
\cite{brady_vicic1995}, and are responsible 
for many physical phenomena, such as the Weissenberg (``rod-climbing'') effect in Couette rheometry 
\cite{weissenberg1947,macosko_book,bird_book1,joseph_fosdick_I_1973,joseph_fosdick_II_1973} 
or the extrudate swell of fluids emerging from a tube \cite{larson1988}.   

The first normal-stress difference $N_1$ is defined for shear flow as the difference between
normal 
stresses in the flow 
and gradient direction, respectively, whereas the second normal-stress difference $N_2$ is given
by the difference 
between normal stresses in the gradient and vorticity (neutral) direction. In Cartesian coordinates
with flow in the $x$ direction and shear gradient in the $y$ direction, this yields $N_1\equiv
\sigma_{xx}\!-\sigma_{yy}$ 
and $N_2\equiv \sigma_{yy}\!-\sigma_{zz}$, where the $\sigma_{ij}$ are stress tensor elements 
\footnote{The sign
convention for both the hydrostatic pressure and the components of the deviatoric stress tensor 
is often a source of confusion. In the present work, we follow 
Refs. \cite{larson1999,wagner_book,macosko_book}, for which the hydrostatic pressure exerted
on a fluid element by its surrounding is negative (compression) and tensile stress is positive.
Some authors prefer to keep both the hydrostatic pressure and the deviatoric
stress components positive in compression (see p. $vii$ of the preface in \cite{bird_book1}).}. 
In experiment, 
the magnitudes of $N_1$ and $N_2$ determine the normal force acting on the plates of a 
rheometer, although the details of this relationship will depend on the geometry of the applied flow 
(e.g., cone-plate, plate-plate) and on boundary conditions \cite{macosko_book}. 
For example, in a cone-plate rheometer, $N_1$ is directly proportional to the force per unit area acting on the plate, which tends to push the plates apart if $N_1>0$, but tends to pull them together if $N_1<0$ \cite{larson1999}. 
The existence of normal stresses can be viewed as a consequence of distortion of the pair correlations 
away from their equilibrium forms.

The importance of normal stresses for the flow of non-Newtonian fluids is most clearly demonstrated by the 
phenomenon of rod climbing \cite{weissenberg1947}, whereby the fluid climbs up 
a rotating shaft, leading to a dramatic distortion of the meniscus profile relative to its 
quiescent form. 
For polymeric systems, this effect is attributed to the existence of a tension along the (circular)
lines of 
flow, which pulls the liquid radially inwards and, consequently, as a result of molecular crowding
in the vicinity of the rod 
surface, upwards against gravity. 
The magnitude of the normal stresses characterizing flow-line tension in polymeric liquids 
(sometimes referred to as `hoop stresses') is often of
comparable magnitude to the shear stresses 
acting in the system.
Experimentally, the rod climbing exhibited by non-Newtonian fluids can be exploited to
characterize nonlinear material properties. In particular, the low-shear-rate limiting values of
the first and second normal-stress coefficients (respectively, $\Psi_1\equiv N_1/\dot{\gamma}^2$ and
$\Psi_2\equiv N_2/\dot{\gamma}^2$, where $\dot{\gamma}$ is the shear rate) can be obtained from observing 
the shape of the meniscus at the rod surface
\cite{macosko_book,joseph1984}. 
This method avoids the experimental difficulties associated with measuring small stress values directly. 
Recently, a promising alternative technique based on active microrheology has
been proposed to simultaneously measure the first and second normal-stress coefficients 
of a complex fluid \cite{squires2010}.

On the basis of existing rheological data for suspensions of repulsive spherical particles, it
appears that these 
systems usually exhibit a value of $N_1$ which is positive and at least a factor of three larger 
than that of $N_2$, with the latter quantity being negative \cite{larson1999}. 
An exception to this rule is found at high-shear rates, where 
theory \cite{bergenholtz_brady2002}, Stokesian dynamics simulations 
\cite{foss_brady2000}, and experiments \cite{lee2006} have all demonstrated that a change of sign of
$N_1$ from a positive to a negative value can arise when the system enters the shear-thickening
regime. 
Although consensus has yet to be reached, it seems likely that this behaviour is connected to the 
formation of lubrication-aggregated colloidal ``hydroclusters'' \cite{brady_bossis1985,cheng2011}. 
A sign change in $N_1$ as a function of rate can also be found in more complex systems, such as 
polymeric liquid crystals \cite{larson1999}, or in attractive emulsions near the glass transition
under shear flow \cite{montesi_prl2004}. In the latter case, the onset of negative $N_1$ coincides
with the formation of rolling cylindrical flocs along the vorticity direction. 
In complete contrast to the above, a purely negative $N_1$ is observed in extended, 
space-spanning networks, such as semiflexible biopolymer gels, regardless of the 
deformation rate \cite{mackintosh2007}.

Understanding the whole rheology of colloidal dispersions from the underlying microscopic mechanisms
within a unique theoretical framework is a formidable task in nonequilibrium statistical
mechanics \cite{brader_review2010}, even for the simplest case of monodisperse spherical particles. 
Indeed, a full description of the many-body dynamics of colloidal
particles in a dispersion under flow should incorporate the complex interplay
between Brownian motion, potential interactions, solvent-mediated hydrodynamic interactions, 
and the geometry of the imposed (time-dependent) flow  \cite{dhont_book}.
Although an all-encompassing constitutive theory is still lacking,
significant progress has been made in recent decades. 
Early attempts employed a fluctuating diffusion equation to calculate the
nonequilibrium static structure factor of dilute charged suspensions under shear, from which zero-shear 
limit expressions for the viscosity and the normal-stress differences were obtained
\cite{ronis1984,ronis1986}. 
An alternative approach, valid at low volume fraction, is to numerically solve the two-particle
Smoluchowski 
equation for the distorted pair correlations, from which the stress tensor
components can be calculated \cite{brady_vicic1995,bergenholtz_brady2002}.

More recently, the \emph{integration through transients} (ITT) approach has been developed 
which enables the derivation of exact generalized Green-Kubo formulas, namely, expressions 
relating average quantities to time integrals over microscopic correlation functions \cite{fuchs_cates2002}. 
Mode-coupling-type approximations to these exact results (ITT-MCT) then lead to closed expressions for the 
macroscopic stress tensor and microscopic time-correlation functions 
\cite{fuchs_cates2002,brader_prl2007,brader_prl2008}. 
The only required input to the ITT-MCT expressions are the volume fraction and static structure factor, 
which serves as proxy for the bare colloidal interaction potential. 

A central feature of this approach is that it captures the nonequilibrium transition between a fluid
and an amorphous solid \footnote{Shear thickening is not captured by
the theory since hydrodynamic interactions are neglected.}. 
When applied to calculate the stress tensor, this theory provides a fully tensorial constitutive 
equation \cite{brader_pre2012}. 
In principle, this makes possible the calculation of the main rheological functions of a colloidal 
dispersion under arbitrary time-dependent flow,
for any imposed interaction potential and volume fraction, either above or below the glass transition. 
In practice, the simultaneous presence of spatial anisotropy and logarithmic time scales hinders 
numerical implementation: full solutions in three spatial dimensions have not yet been 
achieved. Progress has been made in solving the theory for two-dimensional model fluids 
\cite{henrich2009,kruger2011} and the available numerical results show that the ITT-MCT approach 
makes sensible predictions, in qualitative agreement with Brownian dynamics simulation data.

In order to both facilitate a numerical solution and expose the essential physics of the
microscopic 
theory, simplified schematic models have been proposed \cite{fuchs_cates2003,brader_pnas2009}, which 
aim to provide a simpler set of equations with the essential mathematical structure of the microscopic 
theory. Recent applications of this simplified theory have shown that it provides a consistent and
physically robust approach to the phenomenology of glassy
rheology \cite{brader_oscillatory2010,farage2012,step2012,residual_stress2013}. 
However, in resorting to a schematic description of the full theory \cite{brader_prl2008}, one loses
all microscopic spatial information.

In this paper, we start from the fully microscopic, three-dimensional ITT-MCT constitutive 
equation \cite{brader_prl2008} and analyze the normal-stress coefficients, $\Psi_1$ and $\Psi_2$, 
which emerge in the low-shear-rate limit. 
As these coefficients are independent of the shear rate, they represent genuine material functions, 
with a status similar to the familiar zero-shear viscosity.   
By limiting our investigations to the low rate regime, we can extract from the full constitutive
model 
\cite{brader_prl2008} explicit and tractable 
mode-coupling formulas for the normal-stress coefficients, which retain wave-vector dependence 
and require only the volume fraction and static structure factor as input. 
This enables us to investigate the dependence of both $\Psi_1$ and $\Psi_2$ on the details of the 
interparticle interaction. 
As an illustrative example, we focus on a system of colloidal particles interacting via a hard-core
attractive Yukawa (HCAY) potential, which is known to exhibit a reentrant glass transition 
at high volume fractions as a function of the attraction strength \cite{pham_science2002}. 
The calculated normal-stress coefficients then allow us to make predictions regarding the
rod-climbing 
effect in this model system, namely, the dependence of the surface profile on both the 
volume fraction and the strength of the interparticle attraction.

The paper is organized as follows:  In Sec. \ref{itt}, we outline the formal 
integration through transients approach, which leads to an exact generalized Green-Kubo relation 
for the stress tensor. 
In Sec. \ref{normal}, the normal-stress coefficients are discussed in the context of the 
Green-Kubo formalism. 
In Sec. \ref{mct}, we summarize the mode-coupling constitutive equation of \cite{brader_prl2008} 
which approximates the previously developed exact generalized Green-Kubo expressions. 
In Sec. \ref{perturb}, we exploit the constitutive equation to derive formulas for the
low-shear-rate limit of the three main rheological functions (i.e., the viscosity and the first and
second normal-stress coefficients). 
In Sec. \ref{applic}, we apply our theory to investigate the dependence of the normal-stress
coefficients 
on volume fraction and attraction strength for the HCAY system. 
In Sec. \ref{rod_climbing}, we use the calculated $\Psi_1$ and $\Psi_2$ to predict 
the surface profiles which 
would be obtained in a rod-climbing experiment.  
Finally, in Sec. \ref{concl}, we summarize our results and provide an outlook for future work.

%%%%%%%%%%%%%%%%%%%%%%%%%%%%%%%%%%%%%%%%%%%%%%%%%%%%%%%%%%%%%%%%%%%%%%%%%%%%%%%%%%%%%%%%%%%%%%%%%%%%
%%%%%%%%%%%%%%%%%%%%%%%%%%%%%%%%%%%%%%%%%%%%%%%%%%%%%%%%%%%%%%%%%%%%%%%%%%%%%%%%%%%%%%%%%%%%%%%%%%%%

\section{\label{theory}Theory}

\subsection{\label{itt}Integration through transients}
 
We consider a system of spherical colloidal particles, driven into a steady nonequilibrium state 
by an imposed velocity gradient matrix $\kapp$, whose form we initially do not specify. 
%\footnote{In the following, we will only consider steady shear flows, which means
%that neither $\Psi$ nor $\kapp$ have any time-dependence.}. 
For a system of $N$ Brownian particles dispersed in a solvent, interacting through a 
potential $U_N(\{\rr_i\})$ 
(where $\{\rr_i\} \equiv \{\rr_1, \rr_2,\ldots ,\rr_N\}$, with $\rr_i$ indicating the position of the
$i$ th particle), the equation of motion for the probability distribution
$\pdf$ is given by
\begin{equation}\label{smol_eq}
	\frac{\partial\pdf}{\partial t} = \om\,\pdf,
\end{equation}
where $\om$ is the Smoluchowski operator,
\begin{equation}\label{smol_op}
	\om =\sum_{i=1}^{N}\nabla_{i}\cdot\left[D_0\left(\nabla_i-\beta\F_i\right)-
	\kapp\cdot\rr_i\right],
\end{equation}
with $\beta\!=\!(k_\text{B}T)^{-1}$, bare diffusion coefficient $D_0$, and the
direct force $\F_i=-\nabla_i U_N$ acting on particle $i$. 
%and $\nabla_i=(\partial/\partial x_i,\partial/\partial
%y_i,\partial/\partial z_i)$ is the nabla operator acting on the coordinates of particle $i$. 
Many-body hydrodynamic interactions have not been taken into account.
A formal solution of (\ref{smol_eq}) is given by
\begin{equation}\label{smol_sol}
	\Psi(\{\textbf{r}_i\},t\!\rightarrow\!\infty)
	= \pdfe\, \,+\,\, \beta V\!\!\!\int_{0}^{\infty}
	\!\!\!\!\!\!\!ds\,\pdfe \kapp:\siggh e^{s\omadj}\!\!,
\end{equation}
where $\pdfe$ is the equilibrium Boltzmann distribution function,
$\hat{\sigma}_{\alpha\beta}=-(1/V)\sum_{i=1}^{N}F_{\alpha}^{i}r_{\beta}^{i}$, with
$\alpha,\beta=\{x,y,z\}$, are the components of the potential part of
 the stress tensor $\siggh$ \footnote{In a liquid, it is the potential part of the
stress tensor which dominates, whereas in gaseous systems, the kinetic part is the main contribution
to the stress.}, and $V$ is the volume of the system. 
The full contraction is defined as $\textbf{A}:\textbf{B}\equiv\sum_{\alpha,\beta}A_{\alpha\beta}B_{\beta\alpha}$ and 
the adjoint Smoluchowski operator is given by 
\begin{equation}\label{smol_opadj}
	\om^{\dagger} =\sum_{i=1}^{N}\left[D_0\left(\nabla_{i} + \beta\F_i  \right)
	+\rr_i\cdot\kapp^{T}\right]\cdot\nabla_i.
\end{equation}
The solution (\ref{smol_sol}) is the fundamental result of the ITT approach and expresses the
nonequilibrium probability distribution function as an integral over the entire transient flow
history. 

Nonequilibrium averages of any phase-space quantity $f(\{\rr_i\})$ can thus be expressed as
\begin{equation}\label{neq_av}
	\langle f \rangle_{\text{neq}} = \langle f \rangle +
	\beta V\!\!\!\int_{0}^{\infty}\!\!\!ds\,\big\langle
	\kapp\!:\!\siggh \,\,e^{s\omadj}f\big\rangle,
\end{equation}
where $\langle .\rangle_{\text{neq}}$ denotes an average over the nonequilibrium
probability distribution function (\ref{smol_sol}) and $\langle .\rangle$ is a standard
equilibrium average. If we take $f=\siggh$, then (\ref{neq_av}) reads
\begin{equation}\label{gk}
	\sigg\equiv\langle \sigg \rangle_{\text{neq}} = \langle \siggh \rangle +
	\beta V\!\!\!\int_{0}^{\infty}\!\!\!ds\,\big\langle
	\kapp\!:\!\siggh \,\,e^{s\omadj}\siggh\big\rangle,
\end{equation}
which is an exact Green-Kubo-type relation for the stress tensor, expressed as a time integral over
the flow history of the microscopic stress autocorrelation function. 
The term $\langle \siggh \rangle$ yields an isotropic contribution, namely, the interaction-induced
excess 
(over ideal) contribution to the equilibrium pressure, which contributes to neither the
viscosity 
nor the normal-stress coefficients under consideration here.  
%Because the adjoint Smoluchowski operator $\omadj$ depends
%on the velocity gradient matrix $\kapp$, equation (\ref{gk}) is nonlinear in $\kapp$.

\subsection{Normal stress coefficients}\label{normal}
For the special case of steady shear flow, (\ref{gk}) provides a formal result for the 
first normal-stress difference,
\begin{equation}
	N_1=
	\beta V\gam\!\int_{0}^{\infty}\!\!\!dt\,\big\langle
	\hat{\sigma}_{xy}
	\,\,e^{t\omadj}(\hat{\sigma}_{xx}-\hat{\sigma}_{yy})\big\rangle\label{N1},
\end{equation}
where we have used the fact that $\langle \hat{\sigma}_{xx}-\hat{\sigma}_{yy} \rangle=0$.
Using the Taylor expansion
\begin{equation}
	e^{\hat{x}+\alpha\hat{y}} = e^{\hat{x}} +
	\alpha\left[\frac{d}{d\alpha}e^{\hat{x}+\alpha\hat{y}}\right]_{\alpha=0}+\ldots,
\end{equation}
which is valid for arbitrary operators $\hat{x}$ and $\hat{y}$, where $\alpha$ is a scalar
parameter, 
we can expand the right-hand side of (\ref{N1}) to quadratic order in $\dot{\gamma}$, yielding 
\begin{align}
	N_1 &=\beta V\gam\!\int_{0}^{\infty}\!\!\!dt\,\big\langle\hat{\sigma}_{xy}
	\,\,e^{t\omadj_{\text{eq}}}(\hat{\sigma}_{xx}-\hat{\sigma}_{yy})\big\rangle\notag\\
	&\phantom{=}
	+\beta V\gam^2\!\int_{0}^{\infty}\!\!\!dt\,\big\langle\hat{\sigma}_{xy}\left[\frac{d}
	{d\gam}\,\,e^{t\omadj}\right]_{\gam=0}(\hat{\sigma}_{xx}-\hat{\sigma}_{yy})\big\rangle.
	\label{N1_Feyn}
\end{align}
The first term of (\ref{N1_Feyn}) vanishes identically, due to symmetry ($N_1$ is independent 
of the direction of the shear flow). 
Introducing the strain $\gamma=\gam t$, the nonvanishing second 
term in (\ref{N1_Feyn}) can be rewritten as
\begin{equation}
	N_1=\beta V\gam^2\!\int_{0}^{\infty}\!\!\!dt\,t\,\big\langle\hat{\sigma}_{xy}(0)\frac{d}
	{d\gamma}\left[\,\hat{\sigma}_{xx}(t)-\hat{\sigma}_{yy}(t)\right]_{\gamma=0}
	\big\rangle.\label{N1_deformation}
\end{equation}
The infinitesimal strain tensor, $\strain=(\kapp + \kapp^T)t/2$, is a standard deformation measure 
from elasticity theory. In the present case of simple shear, the flow-gradient elements of this
tensor 
are given by $\varepsilon_{xy}=\varepsilon_{yx}=\gamma/2$, such that (\ref{N1_deformation}) 
can be expressed in the alternative form
\begin{equation}
	N_1=\frac{\beta
	V\gam^2}{2}\!\!\!\int_{0}^{\infty}\!\!\!dt\,t\,\big\langle\hat{\sigma}_{xy}(0)\frac{d}
	{d\varepsilon_{xy}}\,\,\left[\hat{\sigma}_{xx}(t)-\hat{\sigma}_{yy}(t)\right]_{
	\varepsilon_{xy}=0}
	\big\rangle.\label{N1eps}
\end{equation}

For a general anisotropic material, Hooke's law can be written as
$\sigma_{ij}=\sum_{k,l}C_{ijkl}\,\,\varepsilon_{lk}$, where the $C_{ijkl}$
are the components of a fourth-order tensor, $\boldsymbol{C}$, called the stiffness or
elasticity tensor, with $i,j,k,l=\{x,y,z\}$. These components are the elastic constants
of the material. 
By analogy with these continuum definitions, we propose to define 
\emph{fluctuating elastic constants}
\begin{equation}\label{fluctuating_constants}
\hat{C}_{ijkl}(t)\equiv
\left(\frac{d\hat{\sigma}_{ij}(t)}{d \varepsilon_{lk}}\right)_{\varepsilon_{lk}=0}.
\end{equation}
Substitution of \eqref{fluctuating_constants} into (\ref{N1eps}) and division of the resulting expression 
by $\dot\gamma^2$ leads to a compact result for the first normal-stress coefficient,
\begin{align}
	\Psi_1 =\frac{\beta
	V}{2}\!\!\!\int_{0}^{\infty}\!\!\!dt\,t\,\big\langle\hat{\sigma}_{xy}(0)
	\left[\hat{C}_{xxxy}(t)-\hat{C}_{yyxy}(t)\right]\big\rangle.\label{N1elastic}
\end{align}
Entirely analogous reasoning leads also to an expression for the second normal-stress coefficient
\begin{align}	
	\Psi_2 = \frac{\beta
	V}{2}\!\!\!\int_{0}^{\infty}\!\!\!dt\,t\,\big\langle\hat{\sigma}_{xy}(0)
	\left[\hat{C}_{yyxy}(t)-\hat{C}_{zzxy}(t)\right]\big\rangle.\label{N2elastic}
\end{align}
Equations \eqref{N1elastic} and \eqref{N2elastic} provide a microscopic interpretation of the 
macroscopic normal-stress coefficients as time integrals over equilibrium correlations between a  
shear stress fluctuation $\hat{\sigma}_{xy}$ and the fluctuating elastic constants. 

The formal expressions \eqref{N1elastic} and \eqref{N2elastic} for the material functions 
$\Psi_1$ and $\Psi_2$ can be compared and contrasted with the standard result for the zero-shear 
viscosity \cite{nagele_bergenholtz1998},
\begin{align}
	\eta_0 = \beta
	V\!\!\!\int_{0}^{\infty}\!\!\!dt\,\big\langle\hat{\sigma}_{xy}(0)\,
	\hat{\sigma}_{xy}(t)\big\rangle.\label{viscosity_gk}
\end{align}
Equations \eqref{N1elastic}--\eqref{viscosity_gk} have in common that the 
correlation function to be integrated involves the fluctuating shear stress element
$\hat{\sigma}_{xy}(0)$, 
which recognizes that the applied flow is a shearing motion in the $x$-$y$ plane. 
The appearance of the fluctuating elastic constants in the correlation functions 
required for \eqref{N1elastic} and \eqref{N2elastic}, as opposed to the simple stress element
$\hat{\sigma}_{xy}(t)$ as in \eqref{viscosity_gk}, expresses the fact that interparticle
interactions 
are responsible for converting shearing motion of the fluid into normal stresses.

For compressible isotropic media, the elastic constants $C_{ijkl}$ are given by 
$C_{ijkl}=\lambda\delta_{ij}\delta_{kl} + \mu(\delta_{ik}\delta_{jl} + \delta_{il}\delta_{jk})$
(see Ref. \cite{landau_elasticity}), which implies that
\begin{equation}\label{C=0}
\begin{aligned}
	\langle\hat{C}_{xxxy}\rangle=0,\\
	\langle\hat{C}_{yyxy}\rangle=0,\\
	\langle\hat{C}_{zzxy}\rangle=0,
\end{aligned}
\end{equation}
are satisfied by the fluctuating elastic constants in equilibrium. 

%The symbols are
%without their hat ($\hat{\phantom{C}}$) because they are macroscopic average quantities. Our
%fluctuating elastic constants can be expressed more precisely:
%\begin{align}
%	\hat{C}_{ijxy}(t) & = \frac{\partial}{\partial\varepsilon_{yx}}\hat{\sigma}_{ij}(t)
%	=\frac{2}{t}\left[\frac{\partial}{\partial\gam}e^{t\omadj}\right]_{\gam=0}\hat{\sigma}_{ij}
%(0) \notag\\
%	& =
%	\frac{2}{\gam t}\int_{0}^{t}\!\!dt' e^{(t-t')\om^{\dagger}_{\text{eq}}}\,\,\delta\omadj\,\, 
%	e^{t'\om^{\dagger}_{\text{eq}}}\,\,\hat{\sigma}_{ij}(0),
%\end{align}
%where in the third equality we used Eq.(\ref{delta_om}) and the Wilcox formula \cite{wilcox1967}
%for the derivative of exponential operators
%\begin{equation}\label{wilcox}
%	\left[\frac{d}{d\alpha}e^{\hat{x}+\alpha\hat{y}}\right]_{\alpha=0}
%	=
%	\int_{0}^{1}d\lambda\,\,e^{(1-\lambda)\hat{x}}\,\,\hat{y}\,\,e^{\lambda\hat{x}},
%\end{equation}
%and we made a change of variable in the integral. According to Eqs.(\ref{C=0}), our definition of
%the fluctuating elastic constants should lead to
%\begin{equation}\label{hatC=0}
%\begin{aligned}
%	\langle\hat{C}_{xxxy}(t)\rangle=0,\\
%	\langle\hat{C}_{yyxy}(t)\rangle=0,\\
%	\langle\hat{C}_{zzxy}(t)\rangle=0.
%\end{aligned}
%\end{equation}

\subsection{\label{mct}The MCT constitutive equation}

The application of MCT-type projection operator methods to approximate the stress autocorrelation
function 
in (\ref{gk}) leads to a closed microscopic constitutive equation for arbitrary steady flow 
\cite{brader_pre2012}
\begin{align}\label{constit}
	\sigg =& \frac{1}{\beta\,32\pi^3}\int_{0}^{\infty}\!\!\!dt\int \!d\kk
	\left[\frac{\partial}{\partial t}\left(\kk\cdot\B(t)\cdot\kk\right)\kk\kk\right]
	\notag\\
	\phantom{=}& \hspace*{2.15cm}\times \left[\left(\frac{S'_{k}S'_{k(t)}}{k\,k(t)S_k^2}\right)
	\Phi_{\kk(t)}^{2}(t)\right],
\end{align}
where $\kk\kk$ is a dyadic product with components 
$(\kk\kk)_{\alpha\beta}=k_{\alpha}k_{\beta}$, and
$S_k$ and $S'_{k}$ are the \emph{equilibrium} static structure factor and its derivative,
respectively. 

The Finger tensor $\B(t)$ is a standard nonlinear deformation measure \cite{doi_edwards}, 
which is defined via the deformation tensor $\E(t)$ according to
\begin{equation}\label{finger}
	\B(t)\equiv \E(t)\cdot\E^{T}(t)=e^{t\kapp}\cdot e^{t\kapp^{T}}.
\end{equation}
The time-dependent
wave vectors in (\ref{constit}) are the reverse-advected wave vectors, $\kk(t)\equiv\kk\cdot\E(t)
=\kk\cdot e^{t\kapp}$, and their presence in the microscopic constitutive equation
(\ref{constit}) is a consequence of translational invariance for spatially homogeneous flows. 
External flow thus enters \eqref{constit} via the Finger tensor as well as the 
(magnitude of the) reverse-advected 
wavevectors, in a nontrivial way. Finally, the function $\Phi_{\kk(t)}(t)$ is the normalized transient density correlator, defined as the equilibrium average 
\begin{equation}\label{correl}
	\Phi_{\kk(t)}(t)\equiv\frac{\langle
	\rho^{\ast}_{\kk(t)}e^{t\cdot\omadj}\rho_{\kk}\rangle}{NS(k)}.
\end{equation}
Mode-coupling-type approximations to this quantity yield a closed equation of motion for the 
density correlator,
\begin{align}\label{eom}
	\dot{\Phi}_{\qq}(t) + \Gamma_{\qq}(t)\left[\Phi_{\qq}(t) + \int_{0}^{t}dt' m_{\qq}(t,t')
	\dot{\Phi}_{\qq}(t')\right] = 0,
\end{align}
where $ \Gamma_{\qq}(t)\equiv D_0\,\bar{q}^2(t)/S_{\bar{q}(t)}$ 
is the initial decay
rate, with $\bar{q}$ being the magnitude of the advected wave vector $\bar{\qq}(t)\equiv\qq\cdot
e^{-t\kapp}$. The memory kernel $m_{\qq}(t,t')$ entering in the equation of motion (\ref{eom}) is
given by
\begin{align}\label{memory}
	m_{\qq}(t,t')=& \frac{\rho}{16\pi^3}\int d\kk
	\frac{S_{\bar{q}(t)}S_{\bar{k}(t')}S_{\bar{p}(t')}}{\bar{q}^2(t')\bar{q}^2(t)}\\
	\phantom{=}&
	\times V_{\qq\kk\pp}(t')V_{\qq\kk\pp}(t)\Phi_{\bar{\kk}(t')}(t,t')\Phi_{\bar{\pp}(t')}(t,t')
	,\notag
\end{align}
where $\pp=\qq - \kk$, and the vertex functions are given by
\begin{equation}\label{vertex}
	V_{\qq\kk\pp}(t)=\bar{\qq}(t)\cdot\left[\bar{\kk}(t) c_{\bar{k}(t)}
	+ \bar{\pp}(t) c_{\bar{p}(t)}  \right],
\end{equation}
with the Ornstein-Zernike direct correlation function $\rho c_{k}=1-(1/S_k)$.

Equations (\ref{constit})--(\ref{vertex}) thus form a closed constitutive theory, where
the only input quantities are the imposed flow $\kapp$ and the static structure factor $S_k$.
Calculation of the latter requires the interaction potential $U_N$ and the volume
fraction of the particles, $\varphi\equiv N(4/3)\pi R^3/V$, with $R$ the radius of a particle.
Although hydrodynamic interactions are absent in the above microscopic description, the
constitutive equation (\ref{constit}) accounts for the competition between the slowing down of the
structural relaxation with increasing volume fraction, which eventually leads to glassy arrest, 
and the shear-induced enhancement of relaxation. 
%The constitutive equation (\ref{constit}) is suitable for describing 
%the rheology of systems ranging from viscous liquids to amorphous solids and provides access 
%to all macroscopic rheological functions. 

%%%%%%%%%%%%%%%%%%%%%%%%%%%%%%%%%%%%%%%%%%%%%%%%%%%%%%%%%%%%%%%%%%%%%%%%%%%%%%%%%%%%%%%%%%%%%%%%%%

\subsection{\label{perturb}Low shear rate expansion and formulas}

The low-shear-rate limit of the stress tensor may be obtained by expanding the
constitutive equation (\ref{constit}) as a power series in $\dot\gamma$. 
%Before proceding with this calculation we emphasize the fact that it is the 
%\emph{equilibrium} static structure factor (evaluated at either the wavevector $k$ or 
%the advected wavevector $k(t)$) and not the nonequilibrium (distorted) one which 
%appears in the microscopic constitutive equation (\ref{constit}). 
% require singular
%perturbation expansions of it (see Ref. \cite{dhont_book}, Section 6.8, for a pedagogical
%explanation, and also Ref. \cite{brady_vicic1995}). 
We henceforth restrict our considerations to a simple shear with flow in the $x$ direction and 
gradient in the $y$ direction, for which the velocity gradient tensor is given by
\begin{equation}\label{flow}
	(\kapp)_{\alpha\beta}=\delta_{x\alpha}\delta_{y\beta}\gam.
\end{equation}
%and remind that the flow intervenes in (\ref{constit}) through both the Finger tensor $\B(t)$ and
%the reverse-advected wavevectors $\kk(t)$ or their magnitude $k(t)$.
Substitution of (\ref{flow}) into (\ref{finger}), and noting that $\kapp^2=\boldsymbol{0}$, yields
the Finger tensor
\begin{equation}
\B(t) =
\left( \begin{array}{ccc}
1+\gam^2 t^2 & \gam t & 0 \\
\gam t & 1 & 0 \\
0 & 0 & 1
\end{array} \right),
\end{equation}
which allows us to directly calculate the time derivative in the first factor of the integrand 
in (\ref{constit}), namely,
\begin{equation}\label{deriv}
	\frac{\partial}{\partial t}\left(\kk\cdot\B(t)\cdot\kk\right)
	=
	2\gam^2 tk_x^2 + 2\gam k_x k_y{.}
\end{equation}
A further source of shear-rate dependence in the integrand of (\ref{constit}) is 
the ratio $S_{k(t)}^{'}/k(t)$. Expansion in $\dot\gamma$ yields
\begin{equation}\label{expansion}
	\frac{S'_{k(t)}}{k(t)} = \frac{S'_{k}}{k} + \gam\frac{d}{d\gam}
	\left(\frac{S'_{k(t)}}{k(t)}\right)\!\!\Bigg|_{\gam=0} + O(\gam^2),
\end{equation}
with
\begin{align}\label{derivative}
	 \frac{d}{d\gam}\left(\frac{S'_{k(t)}}{k(t)}\right)\!\!\Bigg|_{\gam=0}
%	 &=
%	 \frac{d}{dk(t)}\left(\frac{S'_{k(t)}}{k(t)}\right)\!\!\Bigg|_{k(t)=k}\cdot
%	 \frac{dk(t)}{d\gam}\Bigg|_{\gam=0}\notag\\
	 \!\!=
	 \frac{\left(kS''_{k}-S'_{k}\right)}{k^3}k_xk_yt,
\end{align}
where we have used the explicit form of the reverse-advected
wave vector under shear, $\kk(t)=\kk + \gam t k_x$.
Using Eqs. \eqref{deriv}--\eqref{derivative} in (\ref{constit}), and
approximating the density correlator $\Phi_{\kk(t)}(t)$ by its quiescent form $\Phi_{k}(t)$,
yields 
to second order in the shear rate the following expression for the ITT-MCT stress tensor:
\begin{align}\label{main_result}
	\sigg =& \frac{1}{\beta 16\pi^3}\int_{0}^{\infty}dt\int \!\!d\kk \,\,\kk\kk 
	\Bigg\{\gam k_x k_y\left(\frac{S'_k}{S_k}\right)^2
	\frac{1}{k^2}\Phi^{2}_{k}(t)+\notag\\
	\phantom{=}&
	+\,\, \gam^2
	k_x^2\left(\frac{S'_k}{S_k}\right)^2\frac{1}{k^2}\Phi^{2}_{k}(t)\,t \,\,+\notag\\
	\phantom{=}&
	+\,\,\gam^2k_x^2k_y^2\left(\frac{S'_k}{S^2_k}\right)
	\frac{1}{k^4}\left(kS''_k-S'_k\right)\Phi^{2}_{k}(t)\,t \Bigg\} .
\end{align}
%Eq.(\ref{main_result}) is a central result of the present paper. 

Extracting from (\ref{main_result}) the stress components of
interest and integrating in $\kk$ space, we obtain the main three viscometric functions in the 
low-shear-rate limit, namely, the viscosity,
\begin{equation}\label{visc}
	\eta_0 \equiv\frac{\sigma_{xy}}{\gam} =\frac{1}{60\beta 
	\pi^2}\int_{0}^{\infty}\!\!\!dt\int_{0}^{\infty}\!\!\!dk\,\,
	k^4\left(\frac{S'_k}{S_k}\right)^2\Phi_{k}^2(t)\,\,,
\end{equation}
the first normal-stress coefficient,
\begin{align}\label{nsc1}
	\Psi_1 &\equiv \frac{N_1}{\gam^2} \equiv \frac{\sigma_{xx}-\sigma_{yy}}{\gam^2}\notag\\
	&=
	\frac{1}{30\beta\pi^2}\int_{0}^{\infty}\!\!\!dt\int_{0}^{\infty}\!\!\!dk\,\,
	k^4\left(\frac{S'_k}{S_k}\right)^2\Phi_{k}^2(t)\,t\,\,,
\end{align}
and the second normal-stress coefficient,
\begin{align} \label{nsc2}
	\Psi_2 &\equiv \frac{N_2}{\gam^2} \equiv \frac{\sigma_{yy}-\sigma_{zz}}{\gam^2}\notag\\
	&=
	\frac{1}{210\beta\pi^2}\int_{0}^{\infty}\!\!\!dt\int_{0}^{\infty}\!\!\!dk\,\,
	k^4\left(\frac{S'_k}{S^2_k}\right)\left(kS''_k-S'_k\right)
	\Phi_{k}^2(t)\,t\,\,.
\end{align}
Equation (\ref{visc}) is a well-known expression for the zero-shear-rate viscosity \cite{wagner1994,
nagele_bergenholtz1998}, whereas (\ref{nsc1}) and (\ref{nsc2}) are derived here. 
We note that the singular behaviour of the flow distorted structure factor 
\cite{dhont1989} does not play a role, except in the immediate vicinity of the critical 
point.

%---------------------------------------------------------------------------------------------------
\begin{figure}[t]
\begin{center}
\includegraphics[width=0.48\textwidth]{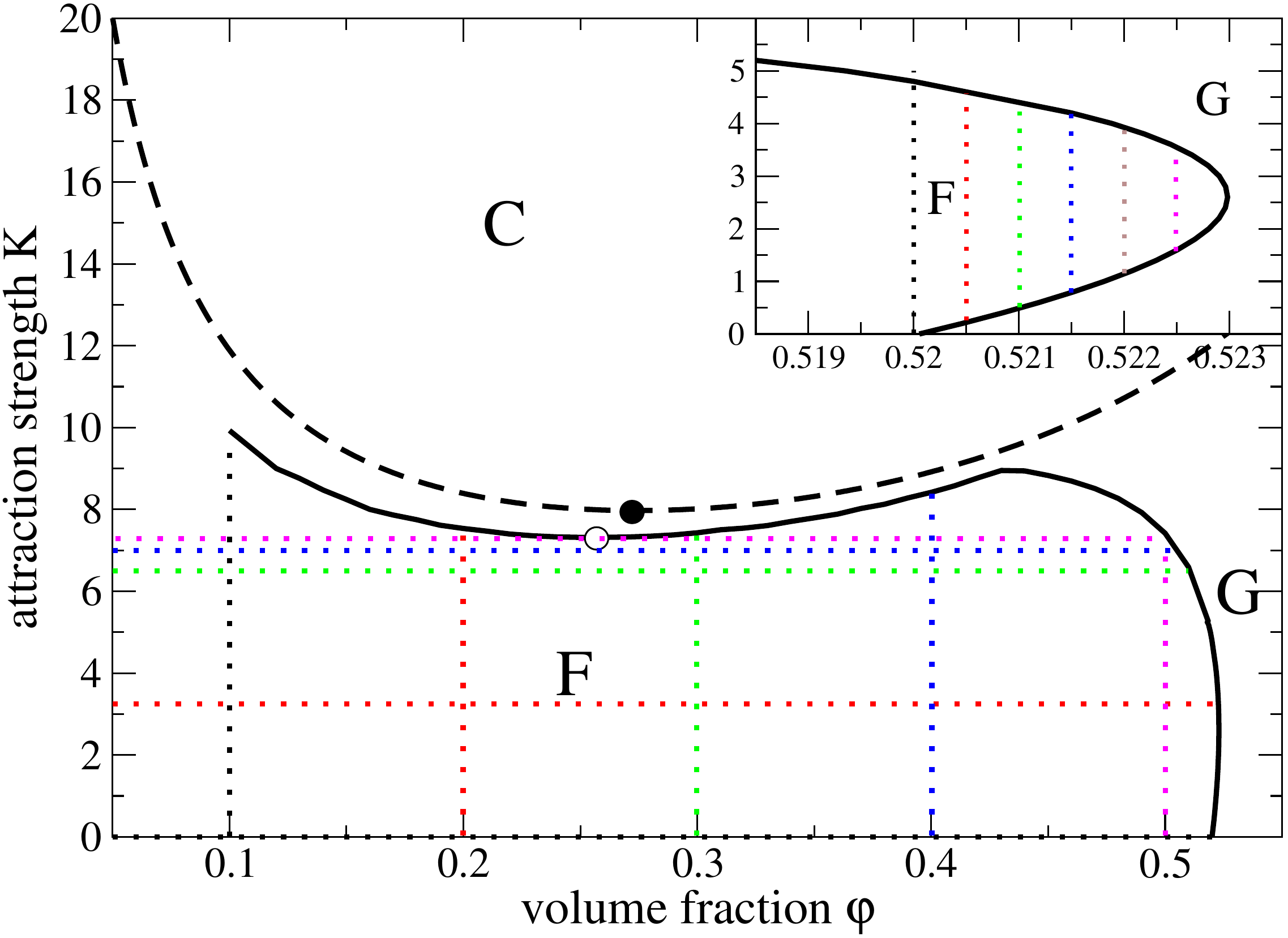}
\caption{Phase diagram for short-range attractive colloids interacting through a HCAY potential
with screening parameter $b=12$. The ordinate axis can also be seen as an inverse temperature axis
($\beta K$). The dashed line is the spinodal and the continuous black
line is the gel-glass line. $C$ indicates the two-phase coexistence region, $F$ is the fluid state,
and
$G$ is the arrested (attractive and repulsive) states. The filled circle ($\bullet$) indicates the
critical point and the empty circle ($\circ$) indicates the minimum of
the ``gel'' line. Horizontal  and
vertical dotted lines indicate
the explored cuts. The inset is a zoom on the reentrance of the glass transition with additional
vertical cuts indicated by the dotted lines. 
}\label{phase_diagram}
\end{center}
\end{figure}
%---------------------------------------------------------------------------------------------------

The key feature of the mode-coupling results (\ref{visc}), (\ref{nsc1}) and
(\ref{nsc2}) is that they enable parameter-free prediction of the most relevant rheological 
quantities for a colloidal system at any volume fraction (below the
glass transition) and subject to an arbitrary interaction potential. 
The accuracy of the predictions will, of course, ultimately depend upon the reliability 
of the approximations employed.   
The quiescent density correlator required as input is isotropic and readily calculable 
using established numerical algorithms, thus avoiding the essential numerical difficulty which  
hinders solution of the full constitutive theory for three dimensional systems.   
As we will demonstrate in Section \ref{applic}, the appropriate signs of the normal stress coefficients 
naturally arise from (\ref{nsc1}) and (\ref{nsc2}), namely $\Psi_1>0$
and $\Psi_2<0$, as well as their expected relative magnitude.

\section{\label{applic}Short-range attractive colloids}
Adding an attractive component to the hard-sphere interaction potential supplements the well-known
first-order crystallization transition by a colloidal liquid-gas transition, 
ending in a critical point. 
If the attraction is of sufficiently short range, then dynamic arrest to either an attractive 
glass or gel state may occur (see \cite{sciortino_tartaglia2005}, and the references 
therein, for a review.). 
Indeed, MCT predictions, later confirmed by simulations and experiments, revealed 
the existence of a reentrant glass transition as a function of attraction strength in dense 
suspensions \cite{pham_science2002}.

The familiar \emph{repulsive} glassy state is obtained by increasing the volume fraction of 
polydisperse hard-sphere particles beyond a critical volume fraction. This leads to an arrested state 
for which the motion of any particle on a distance greater than a few percent of its 
radius is hindered by the neighbouring particles forming a cage around it. 
For systems with an additional short-range attraction, an attractive glass or gel state can be 
reached by reducing the temperature at intermediate ($\varphi > 40\%$) or low volume fraction, 
respectively.

In the remainder of this section, we study the low-shear-rate rheology of a system of colloidal
particles interacting
via a HCAY potential. 
In addition to providing a simple model for describing suspensions found in   
industrial \cite{mezzenga2005} and biological processes \cite{foffi_pre2002}, the fact that the 
phase diagram of the quiescent HCAY model presents both colloidal glass and gel transitions \cite{bergenholtz1999} makes this a system of fundamental interest.

%{\color{blue}
%Therefore, a repulsive
%glass can be first melted by cooling, because openings in the cage are created due to bonding,
%and then, by further cooling, an attractive glassy state can reform, the bonds between particles
%becoming stronger and longer lasting.
%}

%%%%%%%%%%%%%%%%%%%%%%%%%%%%%%%%%%%%%%%%%%%%%%%%%%%%%%%%%%%%%%%%%%%%%%%%%%%%%%%%%%%%%%%%%%%%%%%%%%

\subsection{\label{phase_diagr}Phase diagram}
The HCAY interaction potential between two particles separated by a distance $r$ is given by
\begin{equation}
	\beta u(r) = \left\{ 
	\begin{array}{ll}
		\infty, & 0<r<\sigma\notag\\\\
		-\frac{K}{r/\sigma}e^{-b(r/\sigma -1)}, & \sigma<r,
	\end{array}
	\right.
\end{equation}
where the dimensionless parameter $K$ determines the depth of the attractive well, whereas the
reduced screening parameter $b$ sets the range of the attraction. 
The colloid diameter is denoted by $\sigma$. 
In the present work, we employ the value $b=12$, as this choice generates a phase diagram 
exhibiting all the generic features of the model.

In Fig. \ref{phase_diagram}, we show the equilibrium spinodal and nonequilibrium 
glass-gel transition lines. 
The static structure factor used to calculate the spinodal and as input to our mode-coupling 
approximations was calculated within the mean-spherical closure of the Ornstein-Zernike equation 
\cite{hansen_mcdonald1986}. 
Despite the fact that we consider a monodisperse system, the physics of crystallization 
has no influence on the results to be presented in this work; neither the mean-spherical 
approximation nor the mode-coupling theory are capable of capturing the freezing transition. 
The nonequilibrium phase boundary was obtained using a bisection method, based on repeated numerical 
solution of \eqref{eom}--\eqref{vertex} in the zero-shear-rate limit. To decide whether a
statepoint is fluid or glassy, the long-time limits $\Phi_q(t\rightarrow\infty)$ of the
transient density correlators were determined by solving the corresponding algebraic
equation provided by MCT, and checked for nonzero values (see also the appendix in
\cite{bergenholtz1999}).  
The time dependence of the transient density correlators was
calculated using standard algorithms \cite{voigtmann_thesis} on a wave-vector grid with 250 $k$
values at a grid spacing of 0.3.
Finite differences have been used to approximate the derivatives of the structure
factor. 
The horizontal and vertical dotted lines in Fig. \ref{phase_diagram} indicate paths through the
phase 
diagram along which we display results for the viscosity and normal-stress 
coefficients.

%%%%%%%%%%%%%%%%%%%%%%%%%%%%%%%%%%%%%%%%%%%%%%%%%%%%%%%%%%%%%%%%%%%%%%%%%%%%%%%%%%%%%%%%%%%%%%%%%%%%

\subsection{\label{graphs}Rheological functions: results}

%\subsubsection{\label{horiz_cuts}Horizontal cuts}

%---------------------------------------------------------------------------------------------------
\begin{figure}[t]
\begin{center}
\includegraphics[width=0.42\textwidth]{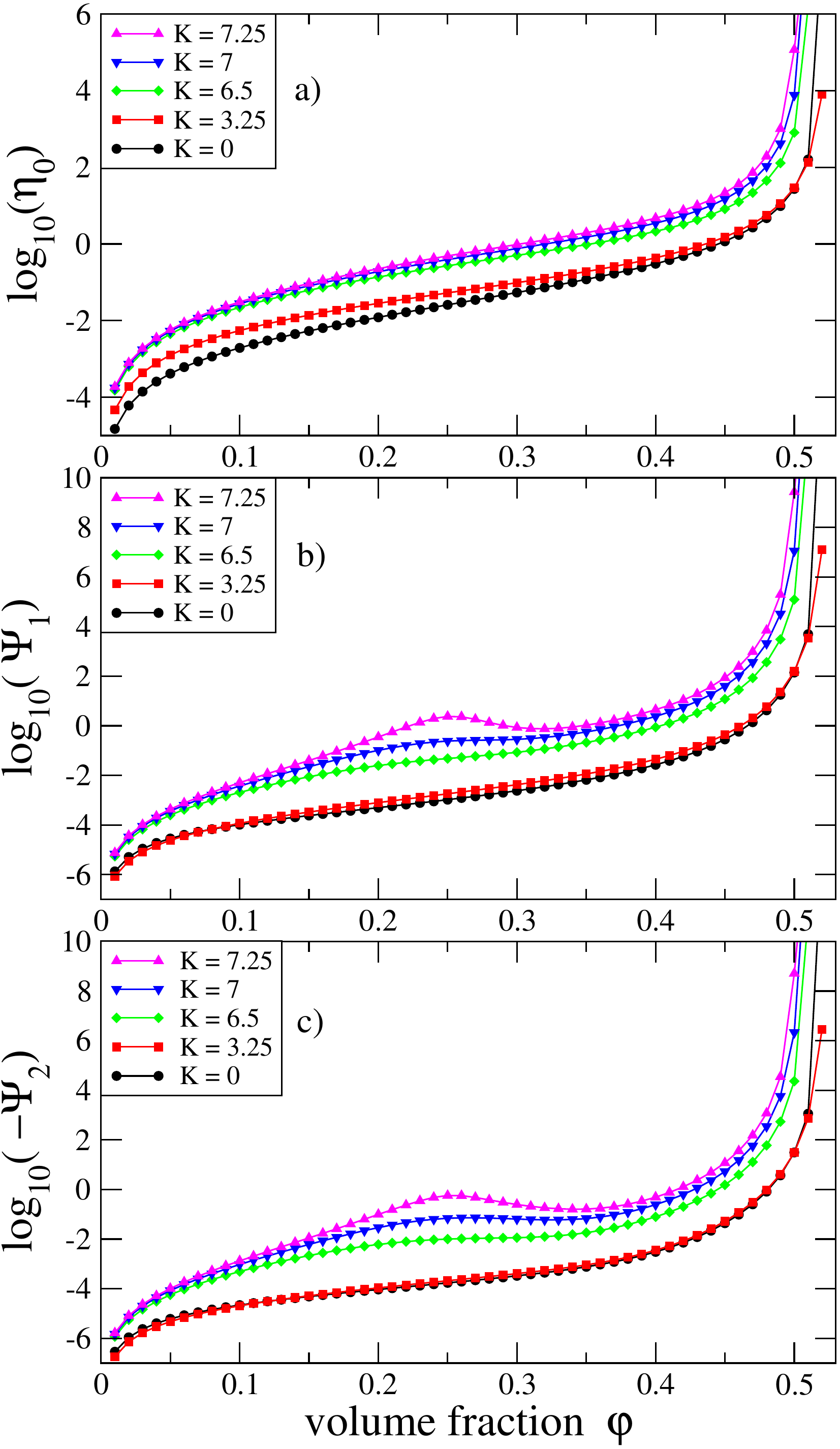}
\caption{Zero-shear-rate limit of (a) the viscosity $\eta_0$, (b) the first normal-stress
coefficient $\Psi_1$, and (c) the second normal-stress coefficient $\Psi_2$ as a function
of the volume fraction $\varphi$, for different values of the potential depth $K$. These
graphs correspond to the paths indicated by the horizontal dotted lines in
Fig. \ref{phase_diagram}.
}
\label{hcuts}
\end{center}
\end{figure}
%---------------------------------------------------------------------------------------------------

Figure \ref{hcuts} shows the volume fraction dependence of the zero-shear viscosity $\eta_0$ and
the 
first and second normal stress coefficients $\Psi_1$ and $\Psi_2$, at different
potential depths $K$ (as indicated in the phase diagram of Fig. \ref{phase_diagram} by the
horizontal
dotted lines).  
Our first observation is that  the $\Psi_2$ predicted by (\ref{nsc2}) is negative for all volume 
fractions, consistent with low-shear-rate experiments and simulations \cite{larson1999}.
A further notable feature of these curves is the influence of the critical point 
%\textcolor{red}{(in
%fact, it is more the influence of the minimum of the gel line!!)} 
on the viscosity and the normal-stress coefficients. 
Whereas the viscosity $\eta_0$ seems to be largely unaffected by the proximity to the
critical point, both $\Psi_1$ and $-\Psi_2$ present a maximum for volume fractions around $0.25$. 
We recall that our present approach does not include the effects of hydrodynamic interactions --
 which are known to have a significant effect upon the low-shear viscosity in the vicinity of 
the critical point \cite{dhont1998} -- and thus provide only the structural contribution. 
The extent of the influence of solvent hydrodynamics on the normal-stress coefficients is, to 
the best of our knowledge, completely unknown, but the present results indicate that the structural 
contribution to these material functions becomes significantly enhanced in the critical region. 
At higher volume fractions, away from the critical point, we expect that hydrodynamic interactions 
will be less important and that the structural component considered here will dominate.  

%---------------------------------------------------------------------------------------------------
\begin{figure}[t]
\begin{center}
\includegraphics[width=0.5\textwidth]{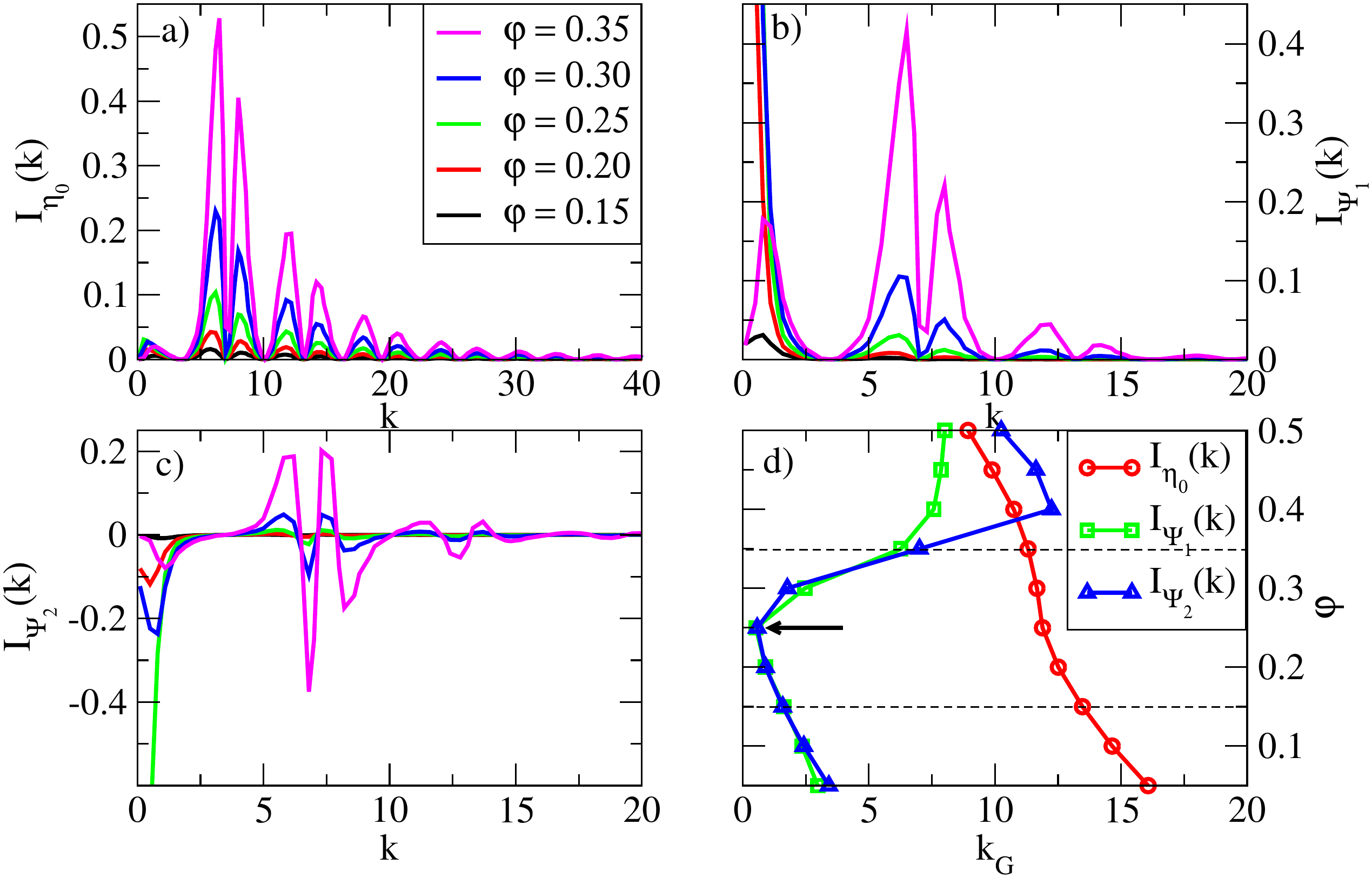}
\caption{Integrands of (a) the viscosity $\eta_0$, (b) the first normal-stress coefficient $\Psi_1$,
and (c) the second normal-stress coefficient $\Psi_2$, for volume fractions $\varphi=\{0.15, 0.20,
0.25, 0.30, 0.35\}$ and an attractive strength $K=7.25$. (d) Plot of the weighted wave vectors
$k_G$ corresponding to the different volume fractions $\varphi$. The horizontal dashed lines
indicate the range of volume fractions considered in (a)--(c), whereas the left arrow
indicates the locus corresponding to the critical point. 
}
\label{integrands}
\end{center}
\end{figure}
%---------------------------------------------------------------------------------------------------
In order to obtain better insight into the microscopic length scales responsible for the 
macroscopic rheological functions shown in Fig. \ref{hcuts}, we show in Fig. \ref{integrands} the 
wave-vector-dependent integrands,
\begin{align}
	I_{\eta_0}(k) & = \frac{1}{60\beta\pi^2}
	k^4\left(\frac{S'_k}{S_k}\right)^2\!\!\int_{0}^{\infty}\!\!\!dt\,\,\Phi_{k}^2(t)\,\,,
	\label{Ivisc}
	\\
	I_{\Psi_1}(k) & =\frac{1}{30\beta\pi^2}\,\,
	k^4\left(\frac{S'_k}{S_k}\right)^2\int_{0}^{\infty}\!\!\!dt\,\,\Phi_{k}^2(t)\,t\,\,,
	\label{Insc1}\\
	I_{\Psi_2}(k) & =\frac{1}{210\beta\pi^2}\,\,
	k^4\left(\frac{S'_k}{S^2_k}\right)\left(kS''_k-S'_k\right)\!\!
	\int_{0}^{\infty}\!\!\!dt\,\,\Phi_{k}^2(t)\,t\,\,,\label{Insc2}
\end{align}
over which we integrate to obtain $\eta_0$, $\Psi_1$ and $\Psi_2$. 
The curves shown in Figs. \ref{integrands}(a)--(c) have been calculated at a fixed potential depth 
($K=7.25$) and for volume fractions over the range $\varphi=0.15 - 0.35$. 
In Fig. \ref{integrands}(d), we show the integral  
$k_G \equiv \int dk\, k I_{\alpha}(k)/ \int dk I_{\alpha}(k)$, where $\alpha=\{\eta_0, \Psi_1,
\Psi_2\}$,  
for the different volume fractions considered. 
This integral measure makes clear the fact that at around $\varphi=0.25$ (namely, close to the
critical point, 
indicated by the arrow in the figure), the wave vectors contributing the most to both
$\Psi_1$ and $\Psi_2$ are at $k_G\approx 0$. 
One can thus conclude that the first and second normal-stress coefficients in the vicinity of the 
critical point are dominated by long-range spatial correlations.    
In contrast, for the viscosity, the value of $k_{G}$ remains at relatively large values,
namely, $8.9<k_G<16.1$, which lie above the main peak (located around $2\pi/\sigma$)
of the static structure factor ($k\approx 7$). 
These findings are consistent with the fact that hydrodynamics interactions can become very important 
in the vicinity of the critical point \cite{dhont1998} because the structural contribution is 
not dominated by long wavelength fluctuations, and supports our implicit assumption that the 
structural component considered here provides the main contribution to the normal-stress
coefficients.

%---------------------------------------------------------------------------------------------------
\begin{figure}[t]
\begin{center}
\includegraphics[width=0.42\textwidth]{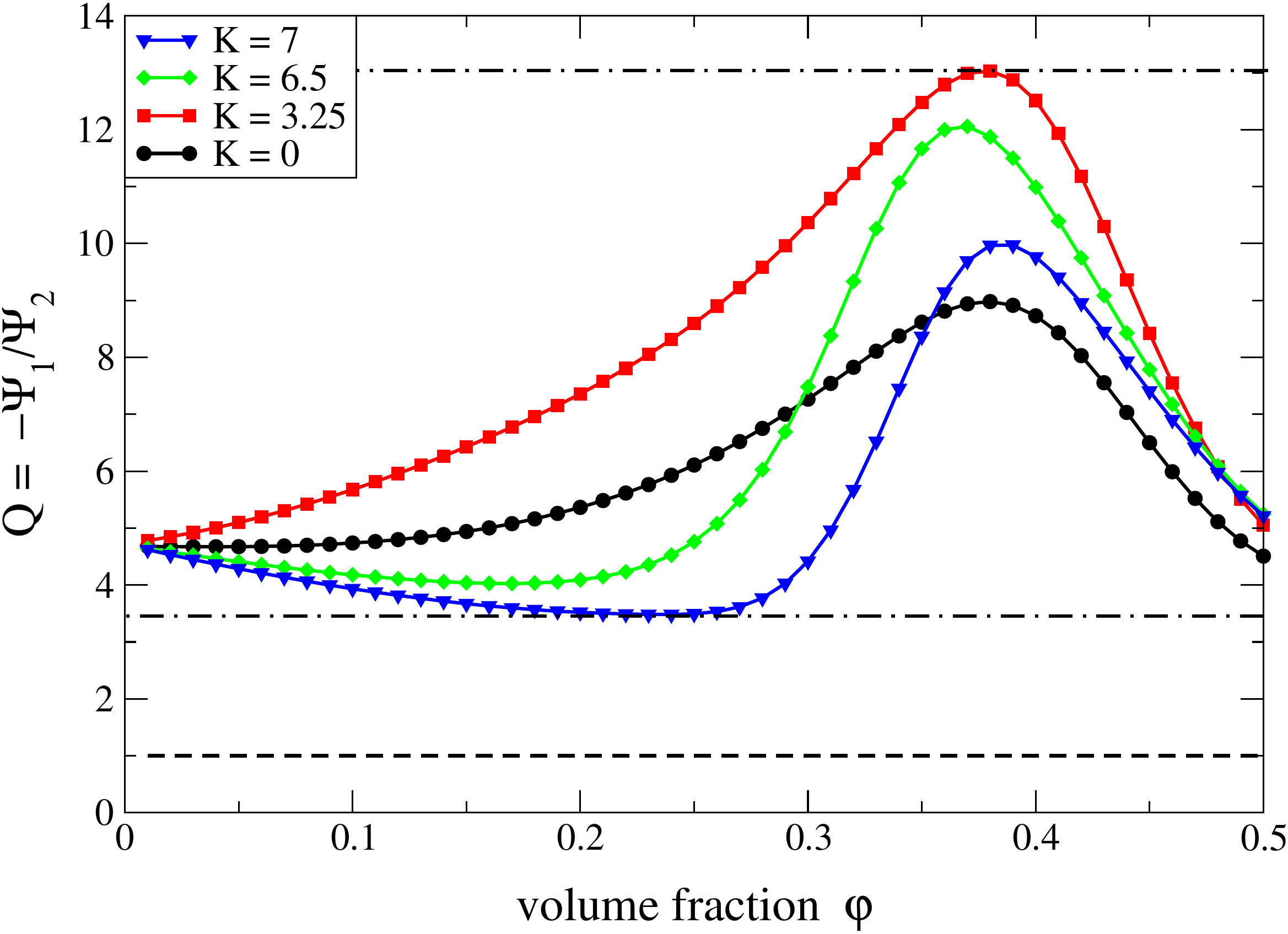}
\caption{Ratio of the normal-stress coefficients ($Q=-\Psi_1/\Psi_2$) as a function of the volume
fraction $\varphi$ at different potential depth values $K$. A ratio $Q=1$ is indicated by the
horizontal dashed line. The other two horizontal dash-dotted lines at $Q\approx3.4$ and
$Q\approx13.0$ indicate the lower and upper boundaries, respectively.
}
\label{Q}
\end{center}
\end{figure}
%---------------------------------------------------------------------------------------------------

We return now to the macroscopic quantities. At first sight, the curves for $\Psi_1$ and $\Psi_2$ 
in Figs. \ref{hcuts}(b) and \ref{hcuts}(c) look qualitatively very similar and it might be expected 
that the ratio $Q\equiv-\Psi_1/\Psi_2$ will not vary as a function of 
volume fraction, for a given potential depth $K$. 
However, as shown in Fig.\ref{Q}, this is not the case and the ratio $Q$ exhibits significant 
structure. 
As mentioned in Sec. \ref{sec:intro}, a lower
boundary to $Q$ of around 3 is to be anticipated on the basis of the available experimental 
data \cite{larson1999}. 
We find that $Q$ remains bounded between approximately $3.4$ and $13.0$ for all volume fractions. 
One of the most striking features of the curves shown in Fig. \ref{Q} is that $Q$ exhibits a 
global maximum at volume fractions $\varphi=0.35-0.4$, for all values of the attraction strength 
investigated, thus indicating the volume fractions for which $\Psi_1$ is numerically most dominant over 
$\Psi_2$. 
This maximum reflects the increasing influence of packing effects and the slowing of structural 
relaxation with increasing volume fraction, although a clear physical interpretation remains 
elusive.
What we observe is that the position (in volume fraction) of the 
global maximum can be correlated with the location of the glass transition boundary shown in 
Fig. \ref{phase_diagram}. 
The influence of the reentrant glass transition is visible when considering the
position of the maxima of $Q$: From $K=0$ to $K=6.5$, the maxima are shifted to lower values of the
volume fraction, whereas for $K>6.5$, they are shifted back to greater values. 
Although calculations performed closer to the critical point are numerically more demanding than 
at other points in the parameter space, the ``bumps'' appearing in Figs. \ref{hcuts}(b) and
\ref{hcuts}(c) in the curves 
approaching the critical point are numerically robust.

%\textcolor{red}{NEW!: We remark that displaying on Fig.\ref{Q} the curve corresponding to $K=7.25$
%would require an extensive numerical study close to the critical point which we renounced to do.
%However, it has to be noticed that the bumps appearing in Fig.\ref{hcuts}b-c in the curves close to
%the critical point are numerically robust and are definitely intrinsic properties of the normal
%stress coefficients.}

%\textcolor{red}{NEW!: }

%---------------------------------------------------------------------------------------------------
\begin{figure}[t]
\begin{center}
\includegraphics[width=0.42\textwidth]{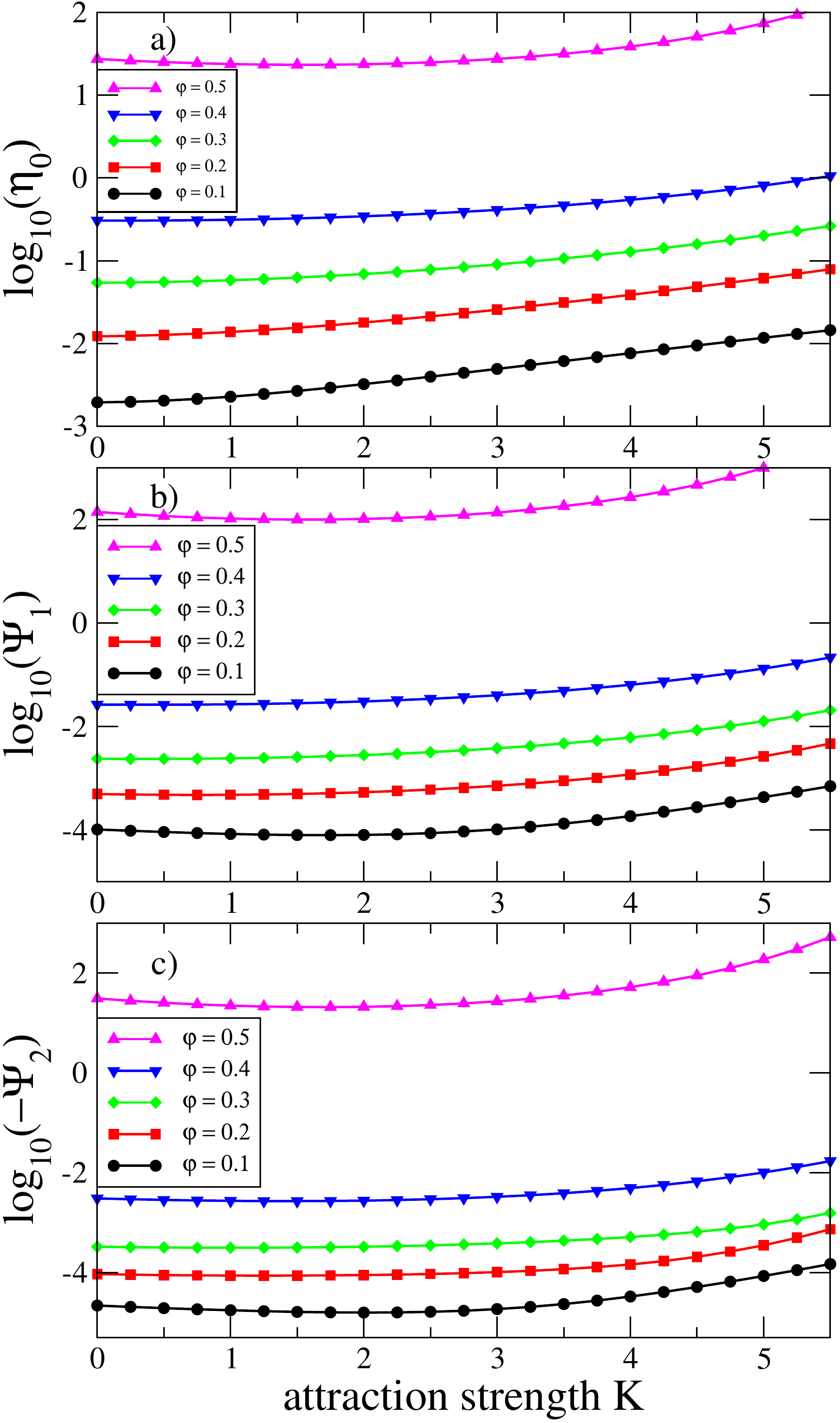}
\caption{Zero-shear-rate limit of (a) the viscosity $\eta_0$, (b) the first normal-stress
coefficient $\Psi_1$, and (c) the second normal-stress coefficient $\Psi_2$ as a function
of the HCAY potential depth $K$, for different values of the volume fraction $\varphi$. These
graphs correspond to the cuts indicated by the vertical dotted lines in
Fig. \ref{phase_diagram}.
}
\label{vcuts}
\end{center}
\end{figure}
%---------------------------------------------------------------------------------------------------
%---------------------------------------------------------------------------------------------------
\begin{figure}[t]
\begin{center}
\includegraphics[width=0.42\textwidth]{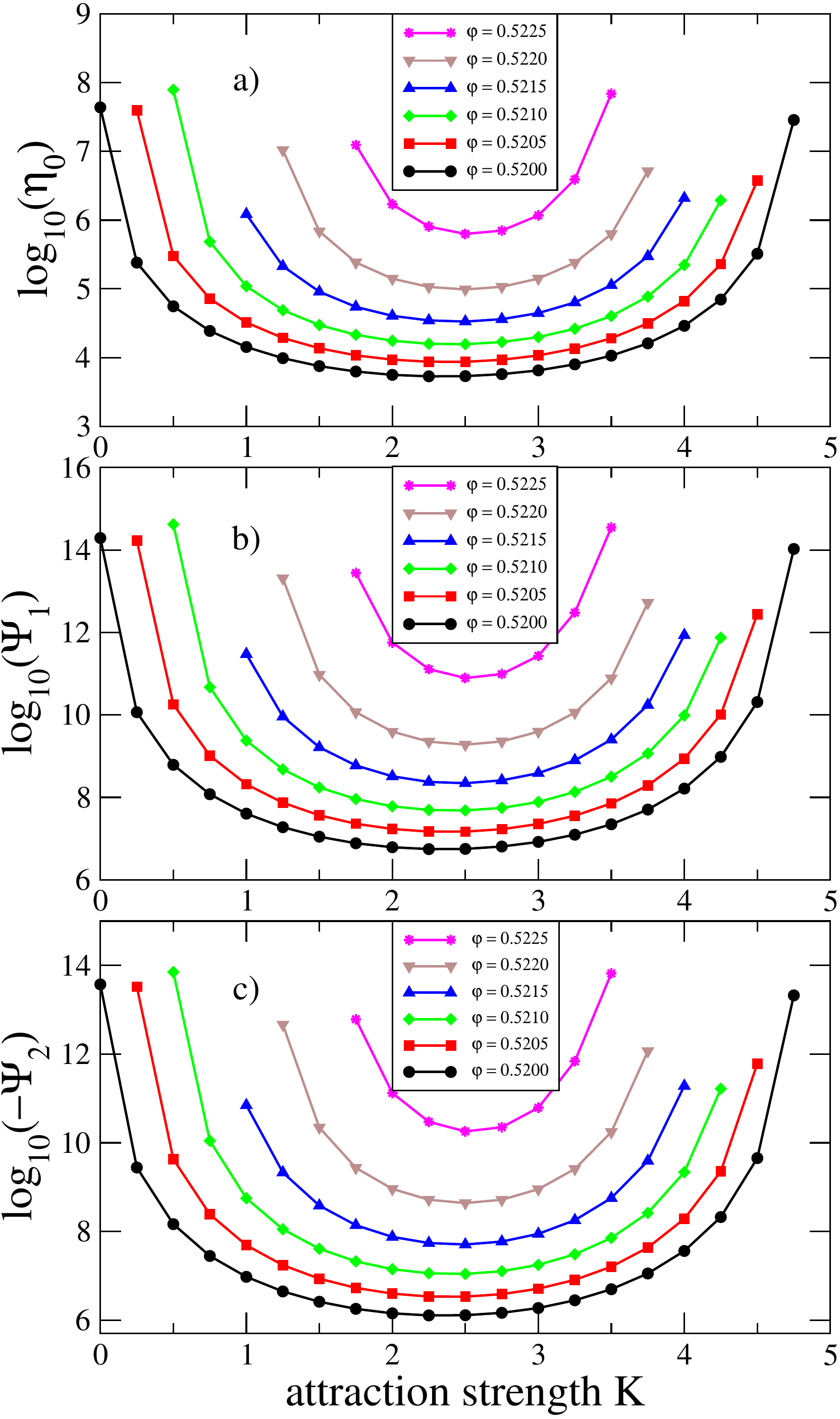}
\caption{Zero-shear-rate limit of (a) the viscosity $\eta_0$, (b) the first normal-stress
coefficient $\Psi_1$, and (c) the second normal-stress coefficient $\Psi_2$ as a function
of the HCAY potential depth $K$, for different values of the volume fraction $\varphi$. These
graphs correspond to the cuts indicated by the vertical dotted lines in the inset of  
Fig. \ref{phase_diagram}.
}
\label{vcutsf}
\end{center}
\end{figure}
%---------------------------------------------------------------------------------------------------

In Figs. \ref{vcuts} and \ref{vcutsf}, we show the viscosity and normal stress coefficients 
as a function of the attraction strength for various values of the volume fraction 
(vertical paths depicted in the phase diagram of Fig. \ref{phase_diagram}). In Fig. \ref{vcuts}(a),
$\eta_0$ develops a minimum at high volume
fraction, whereas in Figs. \ref{vcuts}(b) and \ref{vcuts}(c), both $\Psi_1$ and $\Psi_2$ exhibit
a minimum not only at high volume fraction, but also at a low one ($\varphi=0.1$), which is quite
surprising given that at volume fraction $\varphi=0.1$, the colloidal dispersion is far from
the reentrant region of the phase diagram. 
The nonmonotonic variation of the rheological functions as a function of the attraction strength 
$K$ and the development of a minimum become particularly pronounced in the vicinity of the
reentrant 
glass transition, as is demonstrated in Fig. \ref{vcutsf}. 
In this high volume fraction region, the rheological functions vary by many orders of
magnitude over the range of $K$ values investigated. 
All the minima in Fig. \ref{vcutsf} lie at around $K\approx2.6$, which corresponds to the highest
value of the critical volume fraction (see Fig. \ref{phase_diagram}).

To the best of our knowledge, neither experimental nor simulation data are available yet
for normal-stress coefficients in the case of attractive Brownian particles. 
However, purely repulsive hard-sphere systems have been theoretically investigated and 
numerically simulated \cite{brady_vicic1995, bergenholtz_brady2002, nazockdast2012}. 
For dilute systems at low Peclet number, Brady and Vicic \cite{brady_vicic1995}
predicted normal-stress differences proportional to $\gam^2$, with $N_1>0$ and $N_2<0$, 
clearly consistent with our formulas  (\ref{nsc1}) and (\ref{nsc2}). Moreover they found 
that both $N_1$ and $N_2$ scale with $\varphi^2$, which is exactly the behaviour predicted 
by the present theory (see Fig.\ref{fit}) at $K=0$ and low volume fractions ($0<\varphi<0.15$). 
Concerning the ratio
$Q=-\Psi_1/\Psi_2$, the discrepancy between their value, namely, 1.141, and our value, 4.67, can be
attributed to the different approximations employed in the respective approaches. 
The theoretical predictions made by Nazockdast and Morris \cite{nazockdast2012} at high
volume fractions show that the normal-stress coefficients $\Psi_1$ and $-\Psi_2$ are stronger
functions of
the volume fraction than the zero-shear viscosity $\eta_0$, in agreement with Brady and
Vicic \cite{brady_vicic1995} who predicted $\eta_0/\gam\sim(1-(\varphi/\varphi_m))^{-2}$ and
$\Psi_{1,2}/\gam^2\sim(1-(\varphi/\varphi_m))^{-3}$, where $\varphi_m\approx0.63$ is the random
close-packing volume fraction. We can make the same qualitative statement as Nazockdast and
Morris about the behaviour of our rheological functions. 
Moreover, we find the exponents for the divergence of $\eta_0$, $\Psi_1$, and
$-\Psi_2$ (with the critical volume fraction $\varphi^{\ast}=0.5200527$) to be $-2.55$, $-5.11$, 
and $-5.22$, respectively.

Previous numerical studies of attractive colloidal particles interacting
via square-well or Asakura-Osawa potentials have shown pronounced nonmonotonic behaviour of both the
self-diffusion coefficient \cite{zaccarelli_2002} and the viscosity \cite{puertas2007}. However,
none of the previous works have reported normal-stress coefficients, despite their relevance 
for understanding the rheology of dispersions.

%%%%%%%%%%%%%%%%%%%%%%%%%%%%%%%%%%%%%%%%%%%%%%%%%%%%%%%%%%%%%%%%%%%%%%%%%%%%%%%%%%%%%%%%%%%%%%%%%%%%

\subsection{\label{rod_climbing}The rod-climbing effect}

As mentioned in Sec. \ref{sec:intro}, one striking manifestation of the normal-stress differences in
viscoelastic liquids is the phenomenon of rod climbing (also called the Weissenberg effect) or
rod dipping \cite{wagner_book, macosko_book, weissenberg1947, bird_book1, joseph_fosdick_I_1973,
joseph_fosdick_II_1973}.
Indeed, when a rotating
rod is vertically immersed in a liquid, the latter either climbs or move downwards along the
cylinder because the shearing of the liquid induces stresses both in the gradient and the vorticity
directions. These normal stresses are greater where the shear stress is largest, namely, in the
vicinity of the rotating rod, and since the surface of the liquid is free, the liquid is forced to
move up or down along the cylinder. It is clear that a rotating rod immersed into a
Newtonian liquid will induce a negative surface deflection (or dipping) due to centrifugal forces,
but without additional contributions from the normal stresses, since $N_1=0=N_2$ for Newtonian
liquids.

%---------------------------------------------------------------------------------------------------
\begin{figure}[t]
\begin{center}
\includegraphics[width=0.45\textwidth]{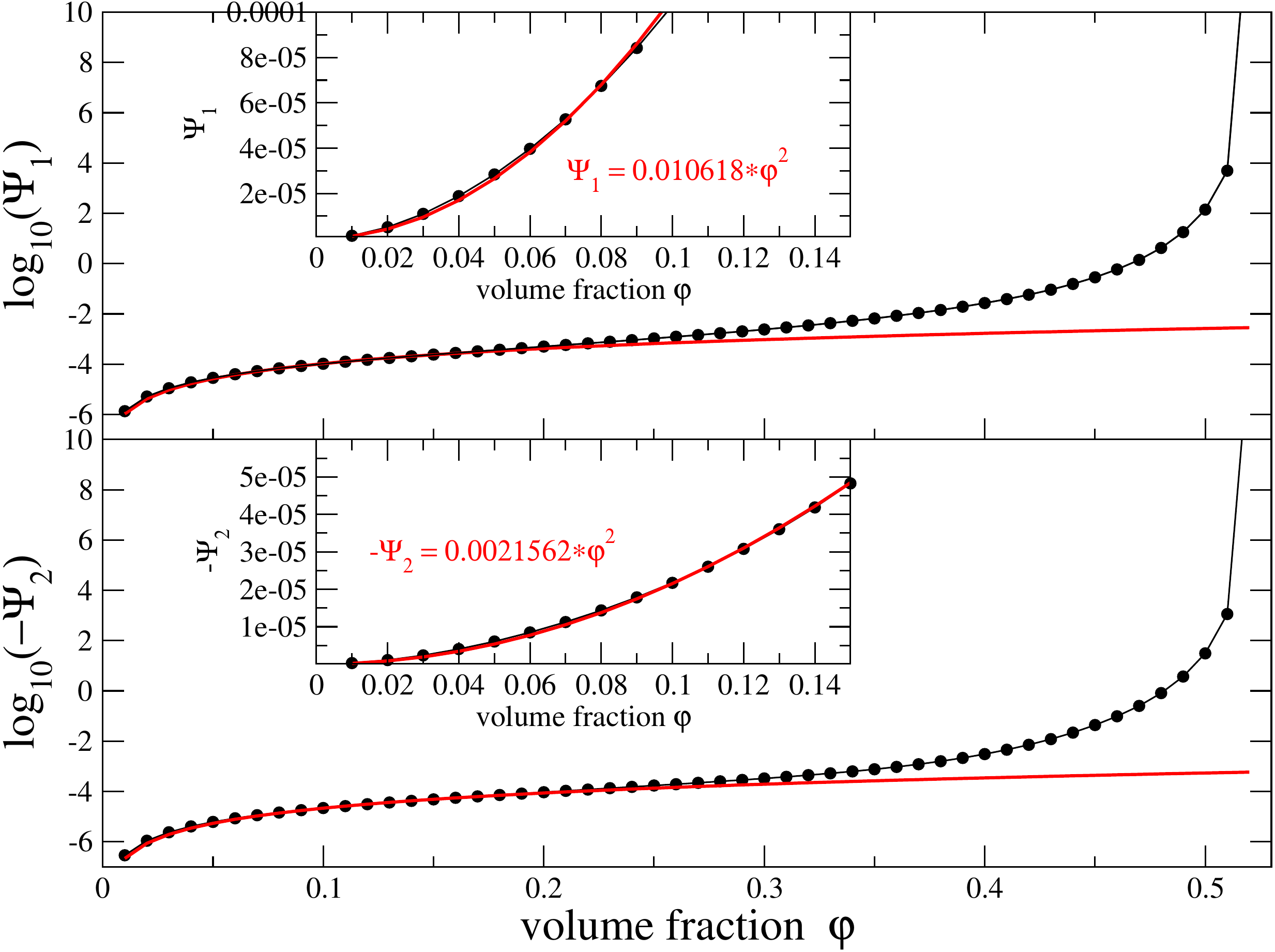}
\caption{First (top) and second (bottom) normal-stress coefficients for pure hard spheres ($K=0$).
The red curves are fitted to the black dots in the range $0<\varphi<0.15$. The insets focus on the
restricted volume fraction range, with linear-linear axes.
}
\label{fit}
\label{fit}
\end{center}
\end{figure}
%---------------------------------------------------------------------------------------------------
Polymer solutions have $N_1>0$ and $N_2<0$, and it is the rod climbing which is
observed in such viscoelastic liquids. On the contrary, dispersions of non-Brownian particles
present both $N_1<0$ and $N_2<0$ (both proportional to $\gam$ rather than to $\gam^2$), and
rod dipping is actually observed \cite{zarraga2000, boyer2011}. 
As we
will see below, the quantity determining whether a viscoelastic liquid will climb or dip
is actually a linear combination of $\Psi_1$ and $\Psi_2$, called the climbing constant and denoted
by $\hat{\beta}$. In the present case of attractive colloidal dispersions, we will see that although
$\Psi_1$ is
always positive and $\Psi_2$ is always negative, the resulting climbing constant $\hat{\beta}$ can
change
its sign from positive (rod climbing) to negative (rod dipping), or vice versa, when
approaching the critical point.

%---------------------------------------------------------------------------------------------------
\begin{figure}[t]
\begin{center}
\includegraphics[width=0.45\textwidth]{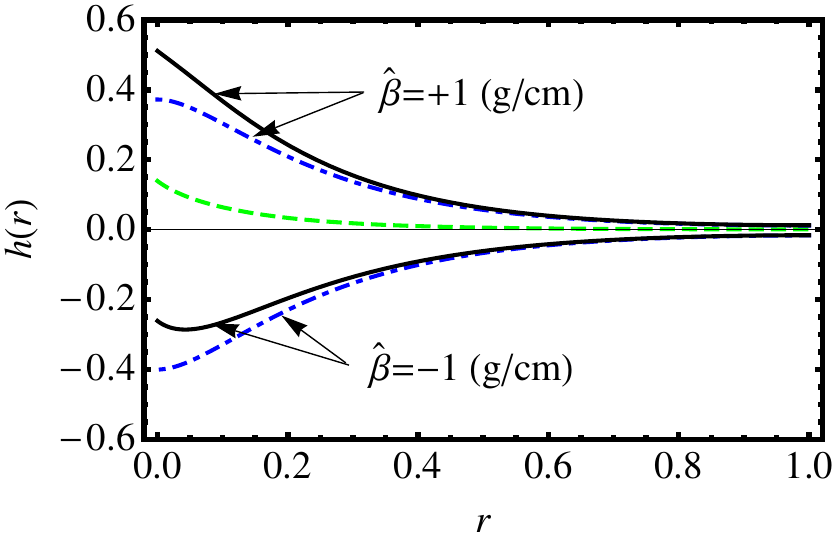}
\caption{Height $h(r)$ (cm) of the free surface of the fluid with respect to the distance $r$ (cm)
from the rod, for two values of the climbing constant $\hat{\beta}$ (g/cm). $r=0$ corresponds to the
rod
surface. The green dashed line is the static deflection $h_s(r)$ due to wetting and the
blue dash-dotted lines correspond to the contribution arising from the rotating rod, namely,
$2\pi^2\omega^2h_2(r)$. The black continuous lines represent the total height of the fluid,
$h(r)=h_s(r)+2\pi^2\omega^2h_2(r)$. Parameters: mass density $\rho=0.88$ g/cm$^3$,
surface 
tension $\mathcal{T}=31$ g$\cdot$s$^{-2}$, rod radius $a=0.32$ cm, contact angle
$m=\tan{55^{\circ}}$, and rotation frequency $\omega = 3.8$ rev/s.}
\label{climbing_dipping}
\end{center}
\end{figure}
%---------------------------------------------------------------------------------------------------

The theory of rod climbing, as well as its application to measurements of the normal-stress
coefficients at low-shear rates, was developed by Joseph and his collaborators
\cite{joseph_fosdick_I_1973, joseph_fosdick_II_1973, joseph1984}. In this theory, the steady
flow profile of a general viscoelastic fluid is given as a perturbation expansion in powers of the
angular velocity $\Omega$ of the rod. The first deviation of the free surface from the static 
profile
(due to wetting) arises at second order $O(\Omega^2)$ and is given by the following
boundary-value problem:
\begin{subequations}\label{BVP_dynamic}
  \begin{gather}
	\frac{\mathcal{T}}{r}\left(rh'_{2}\right)'-\rho g h_2-\frac{\rho a^4}{r^2}+
	\frac{4a^4\hat{\beta}}{r^4}=0,
	\quad a<r<\infty\quad,\label{eq_dynamic}\\
	h'_{2}(a)=0\quad,\label{BC1_dynamic}\\
	(h_2,h'_2)\stackrel{r\rightarrow\infty}{\longrightarrow}(0,0)\quad,\label{BC2_dynamic}
  \end{gather}
\end{subequations}
where $h_2(r)$ is the aforementioned height of the fluid induced by the rotation of the rod, $r$ is
the distance from the center of the rod of radius $a$, $g$ is the gravitational acceleration, and
the dash represents a derivative with respect to $r$. The displacement of the fluid free
surface will generate tensile forces in the surface film. These forces are captured by the first
term of (\ref{eq_dynamic}), where $\mathcal{T}$ is the surface tension. In the absence of surface
tension, the
height $h_2(r)$ of the fluid arises from two distinct contributions. 
One of these is given by the third term in
(\ref{eq_dynamic}) and would exist in the description of a Newtonian fluid: it represents a
depression of the surface due to centrifugal forces, $\rho$ being the mass density of the fluid. 
The other contribution, given by the fourth term of (\ref{eq_dynamic}), accounts for the
non-Newtonian nature of the fluid and describes the climbing (or dipping) along the rod. The
climbing constant $\hat{\beta}$ is defined as $\hat{\beta}\equiv(\Psi_1+ 4\Psi_2 )/2$ and is thus an
instrinsic
property of the fluid \footnote{As mentioned in \cite{joseph_fosdick_II_1973}, it can be shown
that by neglecting wetting, a free surface on the sheared liquid between parallel planes will never
climb. Large radial gradients of the azimuthal shear stress are necessary to induce
climbing.}. A non-Newtonian fluid will climb a rotating rod if $\hat{\beta}>0$. If the surface
tension $\mathcal{T}$
is neglected, it can be easily shown from the modified equation (\ref{eq_dynamic}) that the fluid only
climbs below a critical radius, $r_c=2\sqrt{\hat{\beta}/\rho}$.

%---------------------------------------------------------------------------------------------------
\begin{figure}[t]
\begin{center}
\includegraphics[width=0.49\textwidth]{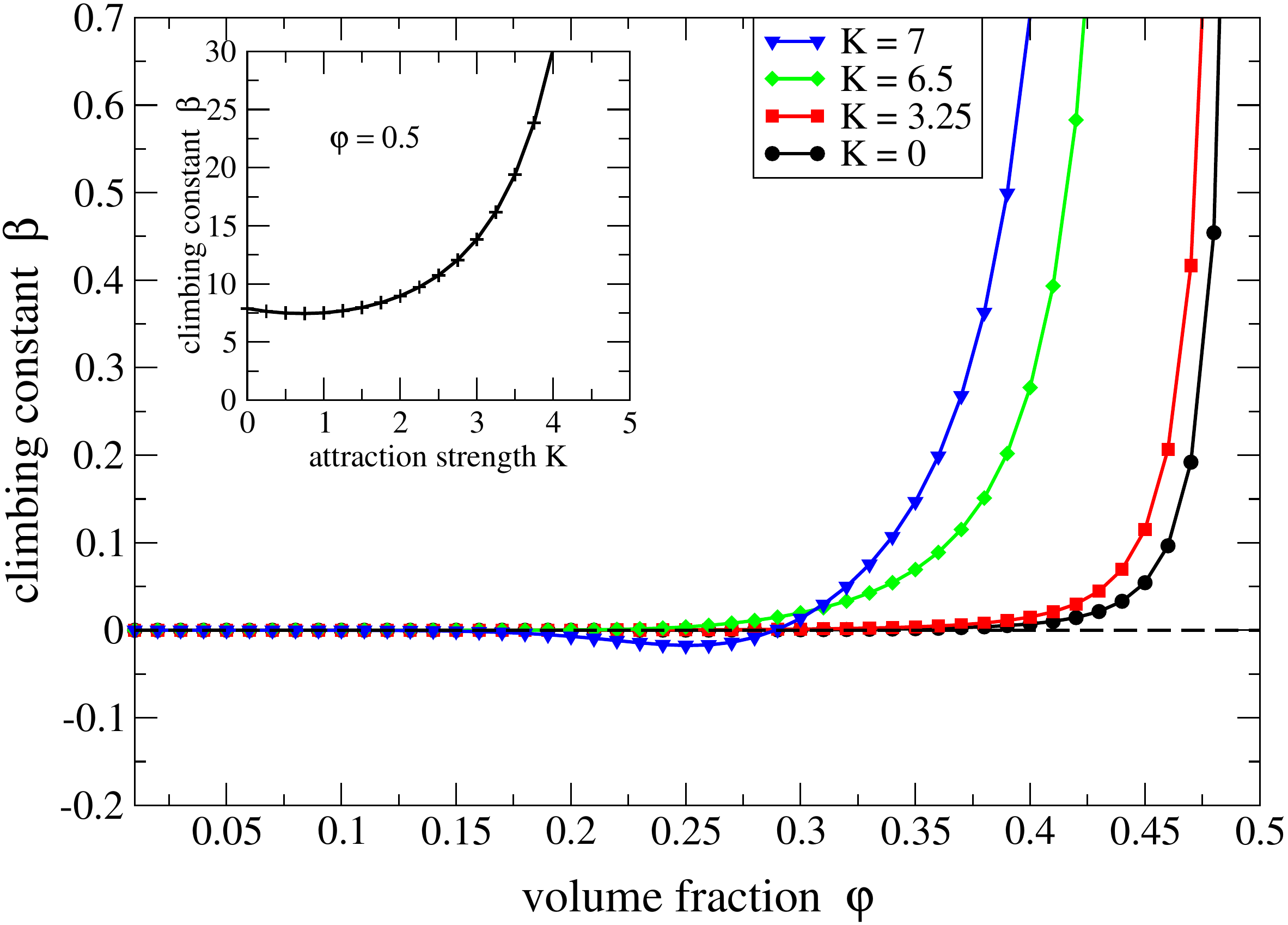}
\caption{Climbing constant $\hat{\beta}$ with respect to the volume fraction $\varphi$ at different
potential depths $K$. The horizontal dashed line indicates $\hat{\beta}=0$. The inset shows the
influence of the glass transition reentrance on $\hat{\beta}$ with respect to $K$, at a volume
fraction close to the glass transition.}
\label{climbing_cst}
\end{center}
\end{figure}
%---------------------------------------------------------------------------------------------------

When the rod is at rest ($\Omega=0$), the static rise $h_s(r)$ of the liquid on the rod due to
wetting is described by the following boundary-value problem:
\begin{subequations}\label{BVP_static}
  \begin{gather}
	\left(\frac{rh'_s}{\sqrt{1+h_s^{'2}}}\right)'-rSh_s=0\quad,\label{eq_static}\\
	h'_s(a)=-m\quad,\label{BC1_static}\\
	(h_s,h'_s)\stackrel{r\rightarrow\infty}{\longrightarrow}(0,0)\quad,\label{BC2_static}
  \end{gather}
\end{subequations}
where $S\equiv\rho g/\mathcal{T}$ and $-m$ is the slope of the free surface at the rod, with 
the contact angle $\alpha$ being defined as $\tan^{-1}(-m)$. In the theory of Joseph
\emph{et al.}, the height of the fluid (with respect to the level of its free surface far from the
rod at rest) is therefore given by the series
\begin{equation}
	h(r;\Omega,m)=h_s(r;m)+h_{2,0}(r)\Omega^2/2+\ldots\label{series}\quad,
\end{equation}
with $h_{2}(r)\equiv h_{2,0}(r)$ and where the leading terms of higher order are
\begin{equation}
	h_{2,1}(r)m\Omega^2/2+h_{4,0}\Omega^4/4!+\ldots\label{higher_order_terms}\quad.
\end{equation}
A good approximation to the expansion (\ref{series}) is given by the truncation
\begin{equation}
	h(r;\Omega,m)\sim h_s(r;m) + h_2(r)\Omega^2/2\label{approx}\quad,
\end{equation}
with this latter being valid while $\omega^4 a<144$, with the frequency $\omega=\Omega/(2\pi)$. If
this
condition is no longer fulfilled, then higher-order terms in the expansion (\ref{series}) must be
considered. In the following, we will present results satisfying the condition $\omega^4 a<144$.
The theoretical surface profiles are thus computed from
\begin{equation}\label{eq_total_height}
	h(r;a,\omega^2,\hat{\beta},m)=h_s(r;a,m) +
	2\pi^2\omega^2h_2(r;a;\hat{\beta})\quad,
\end{equation}
where $h_s(r;a,m)$ and $h_2(r;a;\hat{\beta})$ obey to the boundary-value problems
(\ref{BVP_static}) and (\ref{BVP_dynamic}), respectively. The theory of Joseph {\em et al.} thus 
presents a useful way to determine experimentally the normal-stress coefficients by
identifying the slope of the plot of the total height at the rod surface, $h(a)$, with respect to
the
rotational frequency of the rod. Once this slope is known, the climbing constant
$\hat{\beta}$ can then be calculated from the relation
\begin{equation}
	\frac{dh}{d\omega^2}=\frac{2\pi^2 a}{T\sqrt{S}}\left[\frac{4\hat{\beta}}{4+\lambda}
	-\frac{\rho a^2}{2+\lambda}\right]\quad,
\end{equation}
where $\lambda\equiv a\sqrt{S}$, and finally the normal-stress coefficients can be obtained 
\cite{joseph1984}. 
In Fig. \ref{climbing_dipping}, we attempt to give some feeling for the influence of the 
sign of the climbing constant $\hat{\beta}$ on the free-surface profile.

%---------------------------------------------------------------------------------------------------
\begin{figure}[]
\begin{center}
\includegraphics[width=0.45\textwidth]{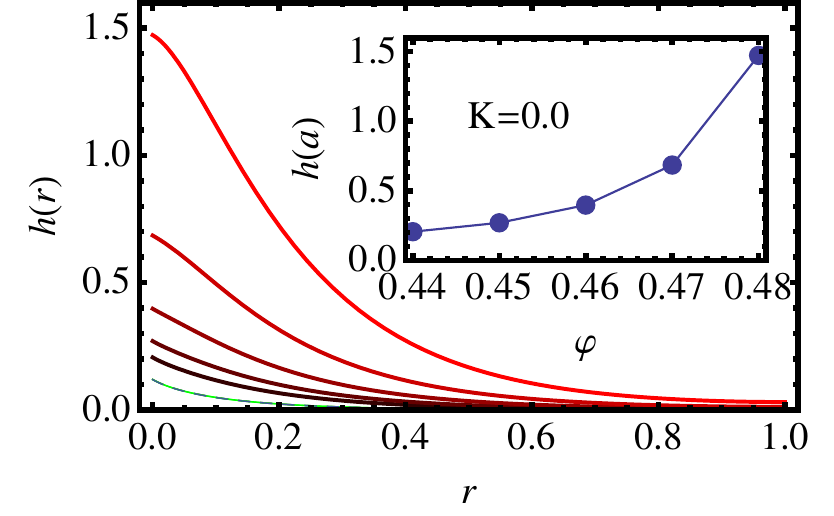}
\caption{Height $h(r)$ (cm) of the free surface of the fluid with respect to the distance $r$ (cm)
from the rod, for a colloidal dispersion of hard spheres at volume fractions ranging from
$\varphi=0.44$ (black lines) to $\varphi=0.48$ (red lines). The green dashed lines indicate the
static
climbing.
Inset: total height at the rod.
See text for chosen parameters.}
\label{climbing_HS}
\end{center}
\end{figure}
%---------------------------------------------------------------------------------------------------

Our derivation of the normal-stress coefficients $\Psi_1$ and $\Psi_2$ at low-shear
rates allows us to calculate the climbing constant $\hat{\beta}$ and make predictions for the
climbing (or dipping) of hard-sphere or attractive colloidal dispersions. 
In Fig. \ref{climbing_cst}, we show the variation of $\hat{\beta}$ with volume fraction 
$\varphi$ for the HCAY system at different attraction
strengths $K$ \footnote{In Fig.\ref{climbing_cst}, SI units are implicitly assumed for
$\hat{\beta}$, namely kg/m.}. 
Although the curves with $K=0$, $K=3.25$, and $K=6.5$ remain monotonic, the one with
$K=7$ displays increased 
structure, due to the proximity to the critical point (see phase diagram of Fig.
\ref{phase_diagram}). We thus predict that a dispersion of hard-sphere or weakly attractive Brownian
particles 
will climb up a rotating rod (although larger volume fractions are required to get significant 
climbing for smaller values of $K$) and that rod dipping will occur when approaching the
critical point. As can be seen in the inset of Fig. \ref{climbing_cst}, at large
volume fractions, e.g., $\varphi=0.5$, the climbing constant $\hat{\beta}$ takes much larger values,
even for $K=0$. Moreover, at these dense values, $\hat{\beta}$ develops a minimum  because of the
influence of the reentrant glass transition and is thus relatively less important at potential
depths around
$K\approx0.75$.

%---------------------------------------------------------------------------------------------------
\begin{figure}[]
\begin{center}
\includegraphics[width=0.45\textwidth]{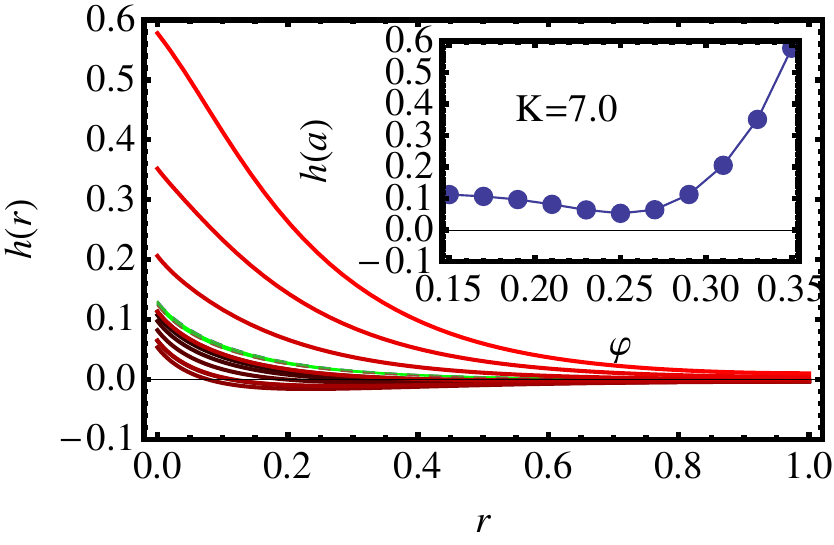}
\caption{Height $h(r)$ (cm) of the free surface of the fluid with respect to the distance $r$ (cm)
from the rod, for a dispersion of short-range attractive colloidal particles interacting via the
HCAY potential. The potential depth is set to $K=7$. Volume fractions range from $\varphi=0.15$
(black lines) to $\varphi=0.35$ (red lines), with a step size of 0.02. The
green dashed lines indicate the static climbing. Inset: total height at the rod.
See text for chosen parameters.
}
\label{climbing_K=7}
\end{center}
\end{figure}
%---------------------------------------------------------------------------------------------------

\begin{figure}[]
\begin{center}
\includegraphics[width=0.45\textwidth]{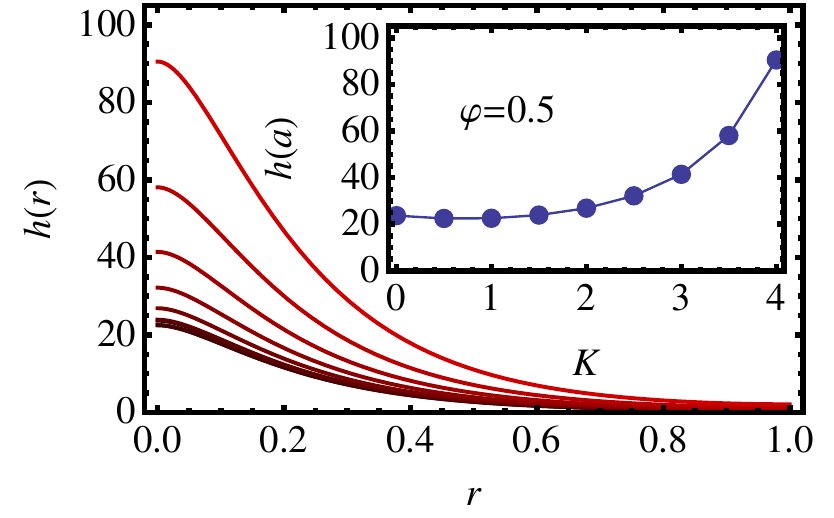}
\caption{Height $h(r)$ (cm) of the free surface of the fluid with respect to the distance $r$ (cm)
from the rod, for a dense dispersion of short-range attractive colloidal particles interacting via
the HCAY potential. The volume fraction $\varphi$ is set to 0.5. The potential depth varies from
$K=0$ (black lines) to $K=4$ (red lines), with a step size of 0.5. Inset: total height at the rod.
See text for chosen parameters.
}
\label{climbing_volfrac_0.5}
\end{center}
\end{figure}

In order to get a better insight regarding the magnitude of the climbing (or dipping) effect, we
choose
a set of realistic values for the different parameters (solvent density $\rho_{\text{solv}}=0.88$
g/cm$^3$ \footnote{\protect\label{solvent}For a dispersion of density matched
colloids in a solvent, the total mass density is given by $\rho=\rho_{\text{solv}}(1+\varphi)$. We
thus choose
$\rho_{\text{solv}}=0.88$ g/cm$^3$.}, surface 
tension $\mathcal{T}=31$ g$\cdot$s$^{-2}$, rod radius $a=0.32$ cm, contact angle
$m=\tan{55^{\circ}}$, and rotation frequency $\omega = 3.8$ rev/s) rather than working with
dimensionless quantities, and we calculate from (\ref{eq_total_height}) the surface profiles for
three different situations depicted in Figs. \eqref{climbing_HS}--\eqref{climbing_volfrac_0.5}. We
point out that the following results for the surface profiles should
be considered as qualitative, rather than quantitative, indications of the physical
phenomenon.

%---------------------------------------------------------------------------------------------------
%---------------------------------------------------------------------------------------------------

Figure \ref{climbing_HS} shows that rod climbing occurs even in the case of hard-sphere colloidal
dispersions, provided that the volume fraction is large enough. With our chosen parameters, the
fluid climbs up to around 1.5 cm at $\varphi =0.48$, which represents about 15 times the height at
the rod due to wetting (see the green dashed curves).
Figure \ref{climbing_K=7} exhibits surface profiles of a semidense dispersion of colloidal
particles
strongly interacting via the HCAY potential with $K=7$. 
Within the range $\varphi\approx0.15 - 0.3$, the climbing constant becomes negative 
(see Fig.\ref{climbing_cst}), which results in rod dipping: the
surface profiles lie below those due to wetting alone. 
This rod-dipping region is induced by the proximity to the critical point. 
For $\varphi>0.3$, the climbing constant becomes positive and
increases monotonically, such that the fluid climbs up the rod.
Finally, Fig.\ref{climbing_volfrac_0.5} shows the surface profiles of a dense colloidal dispersion
($\varphi=0.5$) for different attraction strengths. At such a high volume fraction, the 
elastic component of the dispersion is significant enough to give rise to very strong rod 
climbing.  
Thus, although most of the experiments showing rod climbing have been realized with polymeric 
fluids, the Weissenberg effect is also very prominent in colloidal dispersions. A decisive 
factor in determining the magnitude of the effect is the strength of the elastic contribution 
to the viscoelastic response.

%%%%%%%%%%%%%%%%%%%%%%%%%%%%%%%%%%%%%%%%%%%%%%%%%%%%%%%%%%%%%%%%%%%%%%%%%%%%%%%%%%%%%%%%%%%%%%%%%%%%
%%%%%%%%%%%%%%%%%%%%%%%%%%%%%%%%%%%%%%%%%%%%%%%%%%%%%%%%%%%%%%%%%%%%%%%%%%%%%%%%%%%%%%%%%%%%%%%%%%%%

\section{\label{concl}Conclusion \& outlook}
In this paper, we have shown how the mode-coupling constitutive equation 
\eqref{constit} can be used to develop expressions [Eqs. \eqref{nsc1} and \eqref{nsc2}] for the
first and second normal-stress 
coefficients, opening a path for these important material constants to be calculated from 
first principles. 
Given the system volume fraction and static structure factor, our theory enables us to bridge 
the gap between macroscopic rheological phenomena, such as the rod-climbing effect, and the 
underlying microscopic interactions. 
Although we have neglected the influence of hydrodynamic interactions, we anticipate that 
these will be considerably less important for determining the normal-stress coefficients 
than for the shear viscosity (where hydrodynamic effects are known to be important close to the 
critical point \cite{dhont1998}). 
The theory developed here should thus reliably predict the phenomenology of normal stresses 
and rod climbing, although we anticipate quantitative errors as a result of our various
approximations.  

When our mode-coupling expressions are used as input to the Joseph {\em et al}. theory of 
rod climbing, we can make first-principles predictions for the surface profile of dispersions 
in a Couette rheometer. Qualitative changes in the profile as a function of thermodynamic 
state point can then be investigated in a systematic fashion. This is somewhat contrary to 
the usual experimental practice of determining the interface profile and then using this 
information to infer the normal-stress coefficients. 
It would be of considerable interest to compare our theoretical predictions with data from 
either experiments or simulations on attractive colloids in order to test the qualitative 
trends. Work along these lines is currently in progress. 

In contrast to shear flow, for which $\eta_0$, $\Psi_1$ and 
$\Psi_2$ are all highly relevant, strong flows are characterized entirely by the extensional 
viscosity $\eta_{\text{ext}}\equiv(\sigma_{xx}-\sigma_{yy})/\dot{\epsilon}$ where the stress
components $\sigma_{xx}$ and $\sigma_{yy}$ are those corresponding to an extensional flow whose
extensional strain rate is expressed by $\dot{\epsilon}$. In agreement with the Trouton rules, we
verified that $\eta_{\text{ext}}=3\eta_{0}$, which provides an additional check for the consistency
of the constitutive equation (\ref{constit}) arising from the ITT-MCT formalism.

The present work has addressed rod climbing as a manifestation of normal-stress differences. 
However, these ``hoop stresses'' have also been implicated in the onset of rolling flow in bulk 
and may lie at the origin of vorticity banding \cite{dhont_briels2008}. Whether the present 
theory can predict the onset of such inhomogeneous flow remains a topic for future research.

%%%%%%%%%%%%%%%%%%%%%%%%%%%%%%%%%%%%%%%%%%%%%%%%%%%%%%%%%%%%%%%%%%%%%%%%%%%%%%%%%%%%%%%%%%%%%%%%%%%%
%%%%%%%%%%%%%%%%%%%%%%%%%%%%%%%%%%%%%%%%%%%%%%%%%%%%%%%%%%%%%%%%%%%%%%%%%%%%%%%%%%%%%%%%%%%%%%%%%%%%

\subsection*{Acknowledgements}
We thank Th. Voigtmann for providing numerical code for the solution of the MCT equations. This work
was
supported by the Swiss National Science Foundation.

%%%%%%%%%%%%%%%%%%%%%%%%%%%%%%%%%%%%%%%%%%%%%%%%%%%%%%%%%%%%%%%%%%%%%%%%%%%%%%%%%%%%%%%%%%%%%%%%%%%%
%%%%%%%%%%%%%%%%%%%%%%%%%%%%%%%%%%%%%%%%%%%%%%%%%%%%%%%%%%%%%%%%%%%%%%%%%%%%%%%%%%%%%%%%%%%%%%%%%%%%

%%%%%%%%%%%%%%%%%%%%%%%%%%%%%%%%%%%%%%%%%%%%%%%%%%%%%%%%%%%%%%%%%%%%%%%%%%%%%%%%%%%%%%%%%%%%%%%%%%%%
%%%%%%%%%%%%%%%%%%%%%%%%%%%%%%%%%%%%%%%%%%%%%%%%%%%%%%%%%%%%%%%%%%%%%%%%%%%%%%%%%%%%%%%%%%%%%%%%%%%%
%\bibliography{references.bib}
\bibliography{../../../References/references}
\newpage
\end{document}